\documentstyle[12pt]{article}
\newcommand {\m}{\mu}
\newcommand {\n}{\nu}
\newcommand {\pl}{\partial}
\newcommand {\p} {\phi}

\newcommand {\al}{\alpha}
\newcommand {\be}{\beta}
\newcommand {\ga}{\gamma}
\newcommand {\Ga}{\Gamma}

\newcommand {\la}{\lambda}

\newcommand {\si}{\sigma}
\newcommand {\th}{\theta}

\newcommand {\om}{\omega}

\newcommand {\ep}{\epsilon}

\newcommand {\na}{\nabla}
\newcommand {\del}  {\delta}
\newcommand {\Del}  {\Delta}
\newcommand {\mn}{{\mu\nu}}
\newcommand {\ls}   {{\lambda\sigma}}
\newcommand {\ab}   {{\alpha\beta}}
\newcommand {\half}{ {\frac{1}{2}} }
\newcommand {\fourth} {\frac{1}{4} }
\newcommand {\sqg} {\sqrt{g}}
\newcommand {\fg}  {\sqrt[4]{g}}
\newcommand {\invfg}  {\frac{1}{\sqrt[4]{g}}}

\newcommand {\sqt} {\sqrt{t}}

\newcommand {\Lcal}{{\cal L}}

\newcommand {\Dcal}{{\cal D}}
\newcommand {\Dvec}{{\vec D}}

\newcommand {\Vvec}{{\vec V}}

\newcommand {\xhat}{{\hat x}}

\newcommand {\ptil}{{\tilde \phi}}

\newcommand {\natil} {{\tilde \nabla}}

\newcommand {\intx} {{\int d^2x}}
\newcommand {\intfx} {{\int d^4x}}

\newcommand {\intnw} {{\int d^nw}}

\newcommand {\Achi} {A_{chi}}
\newcommand {\change} {\leftrightarrow}
\newcommand {\ra} {\rightarrow}
\newcommand {\pr}   {{\quad .}}
\newcommand {\com}  {{\quad ,}}
\newcommand {\q}    {\quad}
\newcommand {\qq}   {\quad\quad}
\newcommand {\qqq}   {\quad\quad\quad}

\newcommand {\nn}    {\nonumber}
\newcommand {\vs}[1]  { \vspace*{#1 cm} }

\newcounter{eq}
\newcounter{sc}

\newcommand {\MPL}  {Mod.Phys.Lett.}
\newcommand {\NP}   {Nucl.Phys.}
\newcommand {\PL}   {Phys.Lett.}
\newcommand {\PR}   {Phys.Rev.}
\newcommand {\PRL}   {Phys.Rev.Lett.}

\newcommand {\PTP}  {Prog.Theor.Phys.}
\newcommand {\NC}   {Nuovo Cim.}

\font\smallr=cmr5
\def\ocirc#1{#1^{^{{\hbox{\smallr\llap{o}}}}}}
\def\ogamma{\ocirc{\gamma}{}}

\def\osigma{\ocirc{\sigma}{}}

\def\overleftrightarrow#1{\vbox{\ialign{##\crcr
 $\leftrightarrow$\crcr\noalign{\kern-1pt\nointerlineskip}
 $\hfil\displaystyle{#1}\hfil$\crcr}}}
\def\overnab{{\overleftrightarrow\nabslash}}

\def\tilpsi{{\tilde\psi}}
\def\tbpsi{{\tilde{\bar\psi}}}

\def\Dslash{{}\hbox{\hskip2pt\vtop
 {\baselineskip23pt\hbox{}\vskip-24pt\hbox{/}}
 \hskip-11.5pt $D$}}
\def\nabslash{{}\hbox{\hskip2pt\vtop
 {\baselineskip23pt\hbox{}\vskip-24pt\hbox{/}}
 \hskip-11.5pt $\nabla$}}
\def\xislash{{}\hbox{\hskip2pt\vtop
 {\baselineskip23pt\hbox{}\vskip-24pt\hbox{/}}
 \hskip-11.5pt $\xi$}}
\def\leftnabla{{\overleftarrow\nabla}}

\def\delL{{\delta_{LL}}}
\def\delG{{\delta_{G}}}
\def\delc{{\delta_{cov}}}

\def\gamf{{\ogamma_{(4)}}}
\def\gamt{{\ogamma_{(2)}}}
\def\omef{{\omega_{(4)}}}
\def\omet{{\omega_{(2)}}}
\def\sigf{{\osigma_{(4)}}}
\def\sigt{{\osigma_{(2)}}}

\def\nabslat{{\nabslash_{(2)}}}

\begin{document}
\title{   New Formulation of Anomaly,
          Anomaly Formula and Graphical Representation
          \thanks{US-95-02}
                                 }
\author{
          Shoichi ICHINOSE
          \thanks{ E-mail address:\ ichinose@u-shizuoka-ken.ac.jp\ ;\
Permanent and Mailing address:\ Department of Physics, Universuty of Shizuoka,
          Yada 52-1, Shizuoka 422, Japan}
          and Noriaki IKEDA
          \thanks{ E-mail address:\ nori@kurims.kyoto-u.ac.jp }\\
          Research Institute for Mathematical Sciences, \\
          Kyoto University, Kyoto 606-01,  Japan \\
                          }
\date{  September, 1995 }
\maketitle
\setlength{\baselineskip}{0.54cm}
\begin{abstract}
A general approach to anomaly in quantum field theory is newly formulated by
use of the propagator theory  in solving the
heat-kernel equation. We regard the heat-kernel as a sort of
the point-splitting regularization in the space(-time) manifold.
Fujikawa's general standpoint that the anomalies  come from
the path-integral measure is taken. We obtain some useful formulae
which are valid for general anomaly calculation.
They turn out to be the same as the
1-loop counter-term formulae except some important total derivative terms.
Various anomalies in 2 and 4 dimensional theories
are systematically calculated.
Some important relations between them are concretely shown.
As for the representation of general (global SO(n))
tensors, we introduce a graphical
one . It makes the tensor structure of invariants very transparent
and makes the tensor calculation so simple.
\end{abstract}
\section{Introduction}
A symmetry, in the field theory, imposed at the classical level
sometimes cannot survive
at the quantum level due to the appearance of anomaly. It began to be noticed
around 1950 by Fukuda and Miyamoto
\cite{FM} and was clearly recognized as
the triangle anomaly around 1970
by Adler \cite{A} and Bell and Jackiw \cite{BJ}.
The basic origin at that time was considered as the linear divergence
and its regularization ambiguity
in a triangle diagram. The fresh standpoint on the anomaly was brought
by Fujikawa \cite{F79,F85} around 1980. He regarded the anomaly,
including the Weyl anomaly,
as the Jacobian
in the change of field variable in the path-integral formalism.
In the evaluation Fujikawa paid close attention to the definition
of the path-integral measure.
He also pointed out the connection with the Atiyah-Singer theorem.
The anomaly was generalized to gravitational theories in the general dimension
by Alvarez-Gaum\'{e} and Witten \cite{AW}, especially they found the
pure gravitational anomaly. They exploited the knowledge of differential
geometry
to obtain the explicit form of the chiral U(1) anomaly in 2k dim gravitational
theories. In mid 80's the geometrical aspect of the anomaly
was much examined\cite{AG}.

In the present paper,
special emphasis is on the Weyl anomaly which is different from other
anomalies in the following points.
\begin{enumerate}
\item
Weyl anomaly appears in general theories because it is directly related to
the trace part of the energy-momentum tensor.
It is essentially given by the $\be$-function, which determines
the scaling properties of a theory.
\item
Weyl anomaly, at present, does not seem to be understood
only by the global geometrical (topological) analysis.
This is related to the
fact of item 1.
\item
Quantum effect of gauge or gravitational fields is equally
important to that of matter fields.
\item
The Weyl anomaly is the problem of
the real part of the effective action.
\end{enumerate}
Therefore we can expect the Weyl anomaly contains richer dynamical information
than other anomalies.

Motivated by this expectation, we newly formulate the general anomaly problem.
It is based on ( the coordinate version of )
the {\it propagator approach} in the ordinary perturbative
field theory\cite{BD}.
All explanation is done by the familiar field theory language.
We take the heat-kernel regularization for the ultraviolet divergences.
General formulae of anomalies are obtained.
Many applications are presented.
Weyl anomalies, chiral anomalies(in the flat and curved space), local Lorentz
anomaly and gravitational anomaly are explicitly derived.
Practical usefulness is stressed.

Generally anomaly terms ( especially Weyl anomaly terms),
including their coefficients, are fixed by
calculating (1-loop) quantum fluctuation in the perturbative expansion.
The explicit calculation becomes more and more complicated as we increase
the space(-time) dimension.
In n-dim gravitational theories, we must
treat complicated higher-rank global SO(n) covariants and invariants such as
$\pl_\m\pl_\n h_\ab,\ \pl_\m\pl_\n h_\ls\cdot\pl_\m\pl_\n h_\ls,$\ etc.
 in the weak-gravity expansion:
$g_\ab=\del_\ab +h_\ab,\ |h_\ab|\ll 1,\ (\al,\be,\cdots=1,2,\cdots,n)$.
In order to get rid of the obstacle, we introduce
a graphical representation to treat those terms systematically.
The representation makes
it so easy to list up all independent terms. Although the present paper
deals with $n=2$\ and $4$\ cases only, in order to clearly show the usefulness,
it is applicable to higher-dimensional cases.

The anomaly terms are often compared with the counter-terms. The former
is independent of gauge whereas the latter is not. The former is related
with the consistency of the theory whereas the latter is
related with the renormalization
of the theory. Although both come from the ultra-violet divergences,
their most appropriate regularizations are different:\ the anomaly terms are
commonly
calculated by use of Pauli-Villars regularization (or its variants) whereas
the counter-terms are usually calculated
by use of the dimensional regularization. We will obtain
some direct connection between the two quantities through the anomaly formula
calculated by
the heat-kernel regularization. The formula is very powerful as in the
case of the counter-term formula by 'tHooft\cite{tH}. It is demonstrated
that various anomalies of various theories are derived by the formulae.

The content is organized as follows. In Sec.2 we explain the present
formulation of anomaly taking a simple model of the scalar-gravity
system in n dimension.
It is based on the propagator theory. The heat-kernel regularization
and a graphical representation are taken.
We obtain 2 dim and 4 dim anomaly formula in Sec.3 and 4
respectively. In Sec.5, generalization to fermion-gravity system is presented.
The 2 dim anomaly formula is applied to some interesting anomalies of
2 dim theories in Sec.6. Relations between different anomalies are explained.
In Sec.7 chiral anomalies in flat and gravitational theories are obtained
by the 4 dim anomaly formula. The relation between 2 dim gravitational anomaly
and 4 dim chiral gravitational anomaly is also explained. We conclude in Sec.8.
App.A is prepared for the present notation and some useful formula.
We explain the graphical representation in detail in App.B and C.
App.B is for the global SO(n) covariants and invariants
such as $\pl_\al\pl_\be h_\mn$\ ,
which appear in
concrete substitution of the (weak-field)
expanded terms into the formulae.
App.C is for those which appear as  general terms in the anomaly formulae.

\section{New Formulation of Anomaly and Heat-Kernel Regularization}
\subsection{Anomaly and Jacobian Factor}
Let us explain the present formulation of anomalies taking a simple
example :\ Weyl anomaly in n-dim Euclidean gravity-scalar coupled system.
\begin{eqnarray}
\Lcal[g_\mn,\p]=\sqg (\half g^\mn \pl_\m\p\pl_\n\p+\half qR\p^2)\com\nn\\
q=-\frac{n-2}{4(n-1)}\com\q \label{f.1}
\end{eqnarray}
where $g_\mn$\ and $\p$\ are the metric field and the scalar field.
This Lagrangian is invariant under the local Weyl transformation:
\begin{eqnarray}
g^\mn(x)'=e^{2\al(x)}g^\mn(x)\com\q
\p(x)'=e^{\frac{n-2}{2}\al(x)}\p(x)\com\q
\ptil(x)'=e^{-\al(x)}\ptil(x)
                                                  \com\label{f.2}
\end{eqnarray}
where $\al(x)$ is the parameter of the local Weyl transformation,
$g=det~g_\mn$, and we introduce $\ptil\equiv \fg~\p$ for the
measure $\Dcal \ptil$ to be general coordinate invariant
\cite{F83}.
The partition function ,on the external gravitational field $g_\mn$\ ,
is given by
\begin{eqnarray}
Z[g_\mn]=\int\Dcal\ptil~exp\{~-S[g_\mn,\p]~\}\com\q
S[g_\mn,\p]=\int d^nx\Lcal[g_\mn,\p]\pr  \label{f.3}
\end{eqnarray}
Now let us see the response of $Z[g_\mn]$\ under the Weyl transformation.
\begin{eqnarray}
Z[g_\mn']=\int\Dcal\ptil'exp\{~-\int d^nx~\Lcal[g_\mn',\p']~\}\nn\\
=\int \Dcal\ptil(x)~det\frac{\pl\ptil'(y)}{\pl\ptil(x)}
exp\{~-\int d^nx\Lcal[g_\mn,\p]~\}\pr  \label{f.4}
\end{eqnarray}
Comparing (\ref{f.4}) with (\ref{f.3}), we see the variation of $Z[g_\mn]$\
comes from the Jacobian factor in the Weyl transformation of integration
measure \cite{F79}.
The Jacobian is formally written as
\begin{eqnarray}
J\equiv
det\frac{\pl\ptil'(y)}{\pl\ptil(x)}=det~(e^{-\al(x)}\del^n(x-y)~)\nn\\
=exp(-Tr~[\al(x)\del^n(x-y)]+O(\al^2))\pr\label{f.5}
\end{eqnarray}
In order to regularize the delta function $\del^n(x-y)$,
we introduce the following quantity.
\begin{eqnarray}
G(x,y;t)\equiv <x|e^{-t\Dvec}|y>\com\q t>0\com\nn\\
{\Dvec}_x\equiv \fg(-\na_x^2+qR(x))\invfg\com\label{f.6}
\end{eqnarray}
where \ $t$\ \ will be regarded as a regularization parameter
and is called Schwinger's {\it proper time} \cite{S}.
The operator $\Dvec_x$\
is the hermitian differential (energy) operator which appears in the field
equation for $\ptil$\ :
\begin{eqnarray}
\frac{\del S}{\del \p(x)}=\sqg (-\na^2+qR)\p=\fg{\Dvec}_x\ptil(x)=0\com\nn\\
\Dvec={\Dvec}^{\dag}  \pr                            \label{f.7}
\end{eqnarray}
Here we note the following general features.
\begin{itemize}
\item The physical dimension of the parameter $t$\ is $(\mbox{Length})^2$.
\item The bra- and ket-vector ($<x| ,|y>$) are  rather symbolically
introduced by Schwinger \cite{S}. They
can be more precisely defined as follows\cite{F79} . Let $f_i(x)$ be the
complete ortho-normal eigen-functions of the differential (energy) operator
$\Dvec$\ :
\begin{eqnarray}
\Dvec_xf_i(x)=\la_if_i(x),\ \int d^nx f_i(x)^{\dag} f_j(x)=\del_{ij},\
\sum_i f_i(x)f_i^{\dag} (y)=\del^n(x-y)I\com\label{f.7b}
\end{eqnarray}
where $I$~ is some identity (constant) matrix.
We define the bra and ket vectors, following
Dirac\cite{Di},
by
\begin{eqnarray}
f_i(x)\equiv <x|i>\com\q f_i^{\dag} (x)\equiv <i|x>  \pr\label{f.7c}
\end{eqnarray}
The orthogonality and the completeness conditions (\ref{f.7b}) are realized
by the following requirements:
\begin{eqnarray}
\mbox{Orthonormal Condition}&
<i|j>=\del_{ij}\com\q \int d^nx|x><x|=I\com\nn\\
\mbox{Completeness Condition}&
<x|y>=\del^n(x-y)\com\q \sum_{i}|i><i|=I
\pr\label{f.7d}
\end{eqnarray}
The hermiticity of the operator $\Dvec$\ (or the matrix:\
$D_{ij}\equiv \int d^nx f_i^{\dag} (x)\Dvec_x f_j(x)=\la_i\del_{ij}$\ )
means real eigenvalues:\ $\la_i=\la_i^*$. $G(x,y;t)$\ satisfies
\begin{eqnarray}
G(x,y;t)=<x|e^{-t\Dvec}(\sum_i|i><i|)|y>
=\sum_ie^{-t\la_i}f_i(x)f_i^{\dag} (y)\com
                                                           \nn\\
(\frac{\pl}{\pl t}+\Dvec_x)G(x,y;t)=0\com\q
G(x,y;t)({\overleftarrow {\frac{\pl}{\pl t}}}
        +{\overleftarrow D}^{\dag}_y)=0\com\q             \label{f.7e}
\end{eqnarray}
where ${\overleftarrow D}_x$\ means it operates on the left .
The last two equations show the symmetric property of $G(x,y)$~
with respect to $x$~ and $y$. This property is desirable as the
regularization of $\del^n(x-y)$.
We can regard the equations (\ref{f.7e}) as the precise definition
of $G(x,y;t)$~ introduced in (\ref{f.6}).
\item
{}From the result of the previous item, the positivity of the eigenvalues
$\la_i$\ is required for the well-defined-ness of $G(x,y;t)$~ for
$t\ra +\infty$. (This is important for the evaluation of the effective
action.)
\item
In order for $G(x,y;t)$~ to properly regularize $\del^n(x-y)$~ ,
$\Dvec_x$~ must, at least, satisfy the conditions:\
1)\ hermiticity;\ 2)\ positivity of eigenvalues.
For the correspondence to the anomaly calculation based on
the ordinary (perturbative) quantization, we usually take
, as $\Dvec_x$~, a differential operator which appears in the
field equation. $\Dvec_x$~ is usually an elliptic differential operator.
When the differetial operator in the field equation does not satisfy
the above conditions, ambiguity manifestly appears in the choice of
$\Dvec_x$~ and the anomaly terms are not unique. See Sec.6 for such an
example.
\end{itemize}
In the next subsection, we will see that $G(x,y;t)$\ regularizes
$\del^n(x-y)$
(see eq.(\ref{f.9})).
\subsection{Perturbative Solution of $G(x,y;t)$\ and Heat Equation}
$G(x,y;t)$\ satisfies the following differential equation.
\begin{eqnarray}
(\frac{\pl}{\pl t}+\Dvec_x)G(x,y;t)=0\com\q t>0\pr\label{f.8}
\end{eqnarray}
We solve this equation with the following initial condition.
\begin{eqnarray}
\lim_{t\ra +0}~G(x,y;t)=\del^n(x-y)\pr\label{f.9}
\end{eqnarray}
This equation expresses the regularization of the delta function
$\del^n(x-y)$~where the proper time $t$~plays the role of the regularization
parameter
\cite{AAR,DH}.
For later general use, we write (\ref{f.8}) and (\ref{f.9}) in a more
general form as,
\begin{eqnarray}
(\frac{\pl}{\pl t}\del^{ij}+\Dvec_x^{ij})G^{jk}(x,y;t)=0
                                         \com\q t>0\pr\nn\\
\lim_{t\ra +0}~G^{ij}(x,y;t)=\del^n(x-y)\del^{ij}\com\q
i,j=1,2,\cdots N\com
\label{f.9b}
\end{eqnarray}
where $i$ and $j$ are  the field suffixes , such as
a fermion suffix and a vector suffix. In the present example $N=1$.

For a general theory with the derivative couplings up to the second order,
the operator $\Dvec_x^{ij}$\ can be always written as
\begin{eqnarray}
\Dvec_x^{ij} &=& -\del_\mn\del^{ij}\pl_\m\pl_\n-\Vvec^{ij}(x)\com\nn\\
\Vvec^{ij}(x) &
\equiv & W_\mn^{ij}(x)\pl_\m\pl_\n+N_\m^{ij}\pl_\m+M^{ij}\com\label{f.10}
\end{eqnarray}
where $W_\mn^{ij},\ N_\m^{ij}\ \ \mbox{and}\ \ M^{ij}$\ are external fields
( background coefficient fields ).
In the present example, the above quantities are explicitly written as
\begin{eqnarray}
\Vvec(x) &=& \fg(\na^\mu\na_\m-qR)\invfg-\del_\mn\pl_\m\pl_\n\com\nn\\
W_\mn &=& g^\mn-\del_\mn=-h_\mn+h_{\m\la}h_{\la\n}+O(h^3)\com\label{f.10b}\\
N_\la &=& -g^\mn\Ga^\la_\mn-g^{\la\m}\Ga_\mn^\n
=-\pl_\m h_{\la\m}+O(h^2)\com\nn\\
M &=& -qR+\fourth g^\mn\{\Ga_{\m\la}^\la \Ga_{\n\si}^\si
+2\Ga_{\mn}^\la \Ga_{\la\si}^\si-2\pl_\n \Ga_{\m\la}^\la\}\nn\\
 &=& -q(\pl^2h-\pl_\al\pl_\be h_\ab)-\fourth \pl^2h+O(h^2)\com\nn
\end{eqnarray}
where
\ $g_\mn=\del_\mn+h_\mn\ ,\ h\equiv h_{\m\m}$.
The usage of general coefficients
$W_\mn^{ij},N_\m^{ij}$\ and
$M^{ij}$\ ,instead of their concrete contents,
makes it possible to obtain a general formula for
anomalies\cite{foot1,foot2}
{}.
Let us  solve the differential equation (\ref{f.9b})
perturbatively for the case of weak external fields
($W_\mn^{ij},N_\m^{ij},M^{ij}$). (In the present example, this corresponds
to the perturbation around the {\it flat} space.)
The differential equation (\ref{f.9b}) becomes
\begin{eqnarray}
(\frac{\pl}{\pl t}-\pl^2)G^{ij}(x,y;t) = \Vvec^{ik}(x)G^{kj}(x,y;t)\com\nn\\
\pl^2\equiv \del_\mn\pl_\m\pl_\n=\sum^n_{\mu=1}(\frac{\pl}{\pl x^\m})^2\pr
                                                          \label{f.11}
\end{eqnarray}
In the following we suppress
the field suffixes $i,j,\cdots$ and take the matrix notation.
This equation (\ref{f.11})
is the n-dim heat equation with the small perturbation $\Vvec$.
We prepare two quantities in order to
obtain the solution.
( This approach is popular in the perturbative quantum field theory
under the name of {\it propagator approach}\cite{BD}. In \cite{BD} the momentum
representation is taken, which is to be compared with the coordinate one of
the present approach.
The Weyl anomaly in the string theory was analyzed in this approach
by Alvarez\cite{Al}. )
\begin{flushleft}
i)\ Heat Equation
\end{flushleft}
The  heat equation:
\begin{eqnarray}
(\frac{\pl}{\pl t}-\pl^2)G_0(x,y;t)=0\com\q t>0\label{f.12}
\end{eqnarray}
has the solution
\begin{eqnarray}
G_0(x,y;t)=G_0(x-y;t)=\int\frac{d^nk}{(2\pi)^n}exp\{-k^2t+ik^\m(x-y)^\m\}
                                              ~I_N\nn\\
=\frac{1}{(4\pi t)^{n/2}}e^{-\frac{(x-y)^2}{4t}}~I_N\com\q
k^2\equiv\sum^n_{\m=1}(k^\m)^2\com \label{f.13}
\end{eqnarray}
where $I_N$ is the identity matrix of the size $N\times N$.
$G_0$ satisfies the initial condition:\ $
\lim_{t\ra +0}~G_0(x-y;t)=\del^n(x-y)~I_N
$\ .
We define
\begin{eqnarray}
G_0(x,y;t)=0 \mbox{ for } t\leq 0\pr\label{f.13b}
\end{eqnarray}
\begin{flushleft}
ii)\ Heat Propagator
\end{flushleft}
The heat equation with the delta-function source defines the heat propagator.
\begin{eqnarray}
(\frac{\pl}{\pl t}-\pl^2)S(x,y;t-s)=\del(t-s)\del^n(x-y)~I_N\com\nn\\
S(x,y;t)=S(x-y;t)=\int\frac{d^nk}{(2\pi)^n}\frac{dk^0}{2\pi}
\frac{exp\{-ik^0t+ik\cdot(x-y)\} }{-ik^0+k^2}~I_N  \nn\\
=\th(t)G_0(x-y;t)\com
                                                           \label{f.14}\\
k^2\equiv \sum^\m_{\n=1}k^\m k^\m\com\q
k\cdot x\equiv \sum^n_{\m=1}k^\m x^\m\pr\nn
\end{eqnarray}
$\th(t)$\ is the {\it step function} defined by :
$\th(t)=1\ \mbox{for}\ t>0\ ,\ \th(t)=0\ \mbox{for}\ t<0\ $.
$S(x-y;t)$\ satisfies the initial condition:\ $
\lim_{t\ra+0}S(x-y;t)=\del^n(x-y)~I_N$\
and $
S(x,y;t)=0 \mbox{ for } t\leq 0$\ .

Now the formal solution of (\ref{f.11}) with the initial condition
(\ref{f.9b}) is given by
\begin{eqnarray}
G(x,y;t)=G_0(x-y;t)+\int d^nz\int^\infty_{-\infty}ds~
S(x-z;t-s)\Vvec(z)G(z,y;s)               \pr\label{f.15}
\end{eqnarray}
$G(x,y;t)$\ appears in both sides above.
We can iteratively solve (\ref{f.15}) as
\begin{eqnarray}
G(x,y;t)&=&
G_0(x-y;t)+\int S\Vvec G_0+\int S\Vvec\int S\Vvec G_0+\cdots\com\nn\\
G_1(x,y;t)&\equiv&\int S\Vvec G_0=\int d^nzds S(x-z;t-s)\Vvec(z)G_0(z-y;s)\nn\\
&=&\int d^nz\int^t_0ds G_0(x-z;t-s)\Vvec(z)G_0(z-y;s)\com\label{f.16}\\
G_2(x,y;t)&\equiv&\int S\Vvec\int S\Vvec G_0
=\int d^nz'ds'S(x-z';t-s')\Vvec(z')\nn\\
&&\times\int d^nz ds S(z'-z;s'-s)\Vvec(z)G_0(z-y;s)\nn\\
&=&\int d^nz'\int^t_0ds' G_0(x-z';t-s')\Vvec(z')\nn\\
&&\times\int d^nz\int^{s'}_0ds G_0(z'-z;s'-s)\Vvec(z)G_0(z-y;s)\pr\nn
\end{eqnarray}
Higher-order terms are similarly obtained.
Generally, in n-dim,
the terms up to $G_{n/2}$\ are practically
sufficient for the anomaly calculation\cite{foot3}.
The trace-part of (\ref{f.5})
is given by putting $x=y$\ in the above equations.
\begin{eqnarray}
G(x,x;t) &=& G_0(0;t)+G_1(x,x;t)+G_2(x,x;t)+\cdots\com\q \nn \\
G_0(0;t) &=&\frac{1}{(4\pi t)^{n/2}}~I_N\com\nn\\
G_1(x,x;t)&\equiv&\int \left.S\Vvec G_0\right|_{x=y}      \label{f.17}\\
&=&\frac{1}{(4\pi)^nt^{(n/2)-1}}\int d^nw\int^1_0dr\frac{1}{\{(1-r)r\}^{n/2}}
e^{-\frac{w^2}{4(1-r)}}~\Vvec(x+\sqrt{t}w)e^{-\frac{w^2}{4r}},
                                                 \nn\\
\Vvec(x+\sqrt{t}w)&=&
\frac{1}{t}W_\mn(x+\sqrt{t}w)\frac{\pl}{\pl w^\m}\frac{\pl}{\pl w^\n}
+\frac{1}{\sqrt{t}}N_\m(x+\sqrt{t}w)\frac{\pl}{\pl w^\m}+M(x+\sqrt{t}w)\ ,\nn
\end{eqnarray}
where we introduce some scaled integration variables which are dimension-less:\
$r=\frac{s}{t}\ ,\ w^\m=(z-x)^\m/\sqrt{t}\ $\ . Furthermore
\begin{eqnarray}
&&G_2(x,x;t)\equiv\left.\int S\Vvec\int S\Vvec G_0\right|_{x=y}\nn\\
&=&\frac{1}{(4\pi)^{3n/2}}\frac{1}{t^{(n/2)-2}}\int d^nv \int d^nu
\int^1_0dk\int^k_0dl\frac{1}{\{(1-k)(k-l)l\}^{n/2}}
e^{-\frac{v^2}{4(1-k)}}                    \label{f.18}\\
&&\times \Vvec(x+\sqrt{t}v)e^{-\frac{(v-u)^2}{4(k-l)}}
\Vvec(x+\sqrt{t}u)e^{-\frac{u^2}{4l}}\com\nn
\end{eqnarray}
where $k=s'/t\ ,\ l=s/t\ ,\ v^\m=(z'-x)^\m/\sqrt{t}\ ,\ u^\m=
(z-x)^\m/\sqrt{t}$.

Further analysis will be done for each dimension.
\vs 1
\section{Anomaly Formula in 2 Dimension}
We consider the simplest case of the dimension n=2. From (\ref{f.17}),
we obtain
\begin{eqnarray}
G_0(0;t)=\frac{1}{4\pi t}~I_N\pr\label{2df.1}
\end{eqnarray}
{}From (\ref{f.17}),we obtain
\begin{eqnarray}
G_1(x,x;t) = \frac{1}{(4\pi)^2}\int d^2w\int^1_0dr\frac{1}{(1-r)r}
e^{-\frac{w^2}{4(1-r)}}           \nn\\
\times\{~ (\frac{1}{t}W_\mn(x)+\frac{1}{\sqt}w^\la\pl_\la W_\mn(x)
+\half w^\la w^\si\pl_\la\pl_\si W_\mn(x))
\frac{\pl}{\pl w^\m}\frac{\pl}{\pl w^\n}             \nn\\
+(\frac{1}{\sqrt{t}}N_\n(x)+w^\la\pl_\la N_\m(x))\frac{\pl}{\pl w^\m}
+M(x)~\}~e^{-\frac{w^2}{4r}}+O(t)                    \label{2df.2}\\
=\frac{1}{4\pi}\{~-\frac{1}{2t}W_{\m\m}(x)-\frac{1}{12}\pl^2W_{\m\m}(x)
+\frac{1}{3}\pl_\m\pl_\n W_{\mn}(x)                   \nn\\
-\half\pl_\m N_\m(x)+M(x)~\}+O(t)\com\nn
\end{eqnarray}
where we consider the limit:\ $t\ra +0$\ . The $t^0$-part of
$G_1(x,x;t)$\ , written in terms of $W_\mn,N_\m,M$\ in the final expression of
(\ref{2df.2}),is the anomaly
formula in 2 dimension. The higher-order ones $G_n(x,x;t)\ (n\geq 2)$\
gives higher-terms (with respect to external fields or
$h_\mn$\ ) and is practically
not necessary to fix the anomalies\cite{foot3}.

We apply the above formula to the present example:\ Weyl anomaly
of (\ref{f.1}) with $n=2$.
(Note that $q=0$\ in (\ref{f.1}) for the dimension $n=2$.)
Using the expanded
expression for $W,N,M$\ ,(\ref{f.10b}),we obtain
\begin{eqnarray}
G(x,x;t) &=& G_0(0;t)+G_1(x,x;t)+O(h^2)\nn\\
 &=& \frac{1}{4\pi t}(1+\frac{h}{2})-\frac{1}{24\pi}(\pl^2h-\pl_\m\pl_\n h_\mn)
                                  +O(h^2)+O(t)\label{2df.3}\\
 &=& \frac{1}{4\pi t}(\sqg+O(h^2))-\frac{\sqg}{24\pi}(R+O(h^2))+O(t)\com\nn
\end{eqnarray}
where we use the requirement:\
$\mbox{Tr}[\al(x)\del^2(x-y)]=\lim_{t\ra +0}\lim_{y\ra x}\intx\al(x)G(x,y;t)
$\linebreak
$=\lim_{t\ra +0}\intx \al(x)G(x,x;t)=\mbox{general coordinate invariant}$\ .
(This can be manifestly
shown by the normal coordinate expansion\cite{DeW,Seel,Gil}.)
Therefore the Weyl anomaly is given by
\begin{eqnarray}
\mbox{Weyl Anomaly} &=& \left.\frac{\del}{\del\al(x)}~ln~J\right|_{\al=0}
=\frac{\del}{\del\al(x)}[-\mbox{Tr}~\al(x)\del^2(x-y)]  \nn\\
 &=& - \lim_{t\ra+0}G(x,x;t)
=\sqg [-\frac{1}{4\pi t}+\frac{R}{24\pi}]\pr\label{2df.4}
\end{eqnarray}
The first term of the final form above is divergent and is
renormalized by the cosmological term.

The Weyl anomaly in 2 dim fermion-
gravity system will be considered in Sect.6 where other important
anomalies are also examined.
\section{Anomaly Formula in 4 Dimension }
In this section we will obtain a  formula for anomalies in 4 dim.
\subsection{$G_0(0;t),G_1(x,x;t)$}
{}From (\ref{f.13}) with $n=4$, we obtain
\begin{eqnarray}
G_0(0;t)=\frac{1}{(4\pi t)^2}~I_N\pr \label{4df.0}
\end{eqnarray}
{}From (\ref{f.17}),
\begin{eqnarray}
G_1(x,x;t)=\frac{1}{(4\pi)^4t}\int d^4w\int^1_0dr
\frac{1}{\{(1-r)r\}^2}e^{-\frac{w^2}{4(1-r)}}
\Vvec(x+\sqt w)e^{-\frac{w^2}{4r}}\pr \label{4df.1}
\end{eqnarray}
We notice $t^0$-terms only contribute to the anomalies.
They correspond to log-divergent terms in the effective action:\
$\Ga=\int^\infty_0\frac{dt}{t}G(x,x;t)$.
(\ $t^{-m}$-terms
($m=1,2,\cdots$) contribute to power-divergences,
$t^{+m}$-terms vanish
as $t\ra +0$\ .)
External fields, $W_\mn,N_\m$\ and $M$\ in $\Vvec(x+\sqt w)$\ are
expanded around $t=0$\ as,
\begin{eqnarray}
W_\mn(x+\sqt w) &=& W_\mn(x)+\sqt w^\al\pl_\al W_\mn(x)
+\frac{t}{2}w^\al w^\be\pl_\al\pl_\be W_\mn(x)+\cdots            \nn\\
N_\m(x+\sqt w) &=& N_\m(x)+\sqt w^\al\pl_\al N_\m(x)
+\frac{t}{2}w^\al w^\be\pl_\al\pl_\be N_\m(x)+\cdots    \label{4df.2}\\
M(x+\sqt w) &=& M(x)+\sqt w^\al \pl_\al M(x)
+\frac{t}{2}w^\al w^\be \pl_\al\pl_\be M(x)+\cdots \pr               \nn
\end{eqnarray}
Then we can pick up $t^0$-part of (\ref{4df.1}) as follows.
\begin{eqnarray}
G_1(x,x;t)|_{t^0} &=& \frac{1}{(4\pi)^4}\int d^4w\int^1_0dr
\frac{1}{\{(1-r)r\}^2}e^{-\frac{w^2}{4r(1-r)}}                    \nn\\
&&\times\{~\frac{1}{4!}w^\la w^\si w^\tau w^\om \pl_\la\pl_\si\pl_\tau\pl_\om
W_\mn(x)(-\frac{\del_\mn}{2r}+\frac{w^\m w^\n}{4r^2})             \nn\\
&&+\frac{1}{3!}w^\la w^\si w^\tau \pl_\la\pl_\si\pl_\tau N_\m(x)
(-\frac{w^\m}{2r})+\half w^\la w^\si\pl_\la\pl_\si M(x)~\}     \label{4df.3}\\
&=& \frac{1}{(4\pi)^24!}\{~
-\frac{1}{5}\pl^2\pl^2W_{\m\m}(x)+\frac{6}{5}\pl^2\pl_\m\pl_\n W_\mn (x)\nn\\
&& \qqq -2\pl^2\pl_\m N_\m(x)+4\pl^2 M(x)~\}  \pr                       \nn
\end{eqnarray}
This is one part of the 4 dim anomaly formula. Let us derive the other part.
\subsection{$G_2(x,x;t)$}
{}From (\ref{f.18}), we have
\begin{eqnarray}
G_2(x,x;t)=\frac{1}{(4\pi)^6}\int d^4vd^4u\int^1_0dk\int^k_0dl
\frac{1}{\{(1-k)(k-l)l\}^2}
e^{-\fourth(\frac{v^2}{1-k}+\frac{(v-u)^2}{k-l}+\frac{u^2}{l})}\nn\\
\times\{~
\frac{1}{t}W_\mn(x+\sqt v)(-\frac{\del_\mn}{2(k-l)}
+\frac{(v-u)^\m(v-u)^\n}{4(k-l)^2})                            \nn\\
+\frac{1}{\sqt}N_\m(x+\sqt v)(-\frac{(v-u)^\m}{2(k-l)})+M(x+\sqt v)
{}~\}
                                                            \label{4df.4}\\
\times\{~
\frac{1}{t}W_\ls(x+\sqt u)(-\frac{\del_\ls}{2l}
+\frac{u^\la u^\si}{4l^2})
+\frac{1}{\sqt}N_\la(x+\sqt u)(-\frac{u^\la}{2l})+M(x+\sqt u)
{}~\}\pr
                                                                  \nn
\end{eqnarray}

For the comparison with the (1-loop) counter-term formula, we first consider
the case of 'flat' space:\ $W_\mn =0$\ . In this case the $t^0$-part
of (\ref{4df.4}) is given as
\begin{eqnarray}
G_2^{flat}(x,x;t)|_{t^0} &=& \frac{1}{(4\pi)^2}
[\ -\frac{1}{12}\pl_\n(\pl_\n N_\m\cdot N_\m)     \nn\\
&& +\frac{1}{48}(\pl_\m N_\n-\pl_\n N_\m)^2+\half (M-\half \pl_\m N_\m)^2\ ]
                                 \pr\label{4df.4b}
\end{eqnarray}
Combining the above result and (\ref{4df.3}) for $W_\mn=0$~ ,
$G^{flat}(x,x;t)|_{t^0}$~ is expressed, up to the present approximation, as
\begin{eqnarray}
G^{flat}(x,x;t)|_{t^0}=\frac{1}{(4\pi)^2}
[\
\frac{1}{6}\pl^2(M-\half\pl_\m N_\m)
-\frac{1}{12}\pl_\n(\pl_\n N_\m\cdot N_\m)\ ]\nn\\
+\Del\Lcal_{tH}+O((N,M)^3)\com     \nn\\
\Del\Lcal_{tH}\equiv\frac{1}{(4\pi)^2}[\
\frac{1}{48}(\pl_\m N_\n-\pl_\n N_\m)^2+\half (M-\half \pl_\m N_\m)^2\ ]
                                 \com\label{4df.4c}
\end{eqnarray}
where $\Del\Lcal_{tH}$~ is 'tHooft's 1-loop counter-term formula \cite{tH}
at the present approximation.
This anomaly formula should be compared with
the counter-term formula in the following points:\
1)\ The total derivative terms
have meaning in the anomaly formula;\
2)\ Because the present approximation is
weak-field expansion (of the 1-loop part) up to $G_2$,
the cubic and
quartic terms with respect to the external fields ($N_\m,M$) do not appear
and will appear in $G_3$~ and $G_4$~;\
3)\ The linear terms appear as total derivatives;\
4)\ Symmetries, with respect to interchange of suffixes, are not assumed
for the external fields $N_\m$~ and $M$;\
5)\ The counter-term formula is obtained using the dimensional regularization,
 whereas the present anomaly formula is obtained using the heat-kernel
regularization.
Although
conformal anomalies were discussed  in connection with the
1-loop counter-term formula\cite{HT}, such a direct relation as above
has not been known so far.
The above formula (\ref{4df.4b}) will be used in Sect.7.1

For the anomaly calculation of gravitational theories, we must consider
the general case of $W_\mn$. From the dimensional counting
([$W_\mn$]=(Mass)$^0$\ ,\ [$N_\m$]=(Mass)$^1$\ ,\ [$M$]=(Mass)$^2$\ ,\
[$\pl_\m$]=(Mass)$^1$\ ),
we see the $t^0$-part of $G_2(x,x;t)$\ has three types of terms:\
1)\ $W\times (\pl\pl\pl\pl W,\ \pl\pl\pl N,\ \pl\pl M)$\ ;\
2)\ $(\pl W,N)\times (\pl\pl\pl W,\pl\pl N,\pl M)$\ ;\
3)\ $(\pl\pl W,\pl N,M)\times (\pl\pl W,\pl N,M)$.
Among
them, the most useful ones are type 3) terms, i.e. those terms
which are composed only of (Mass)$^2$-dimensional quantities:\
$(\pl\pl W,\pl N,M)$\ ,because they are sufficient to determine
all anomaly terms\cite{foot3}.
They are given by
\begin{eqnarray}
G_2(x,x;t)|_{t^0}^{(\pl\pl W,\pl N,M)}
=\frac{1}{(4\pi)^6}\int d^4vd^4u\int^1_0dk\int^k_0dl
\frac{1}{\{(1-k)(k-l)l\}^2}                                    \nn\\
\times e^{-\fourth(\frac{v^2}{1-k}+\frac{(v-u)^2}{k-l}+\frac{u^2}{l})}
\{~
\half \pl_\al\pl_\be W_\mn(x)\cdot v^\al v^\be (-\frac{\del_\mn}{2(k-l)}
+\frac{(v-u)^\m(v-u)^\n}{4(k-l)^2})                            \nn\\
+\pl_\al N_\m(x)\cdot v^\al (-\frac{(v-u)^\m}{2(k-l)})+M(x)
{}~\}
                                                            \label{4df.5}\\
\times\{~
\frac{1}{2}\pl_\ga\pl_\del W_\ls(x)\cdot u^\ga u^\del (-\frac{\del_\ls}{2l}
+\frac{u^\la u^\si}{4l^2})
+\pl_\ga N_\la(x)\cdot u^\ga (-\frac{u^\la}{2l})+M(x)
{}~\}\pr
                                                                  \nn
\end{eqnarray}
Other types, 1) and 2),
are also similarly evaluated, but practically they are not necessary
for the anomaly calculation.
\subsection{Graphical Representation of Anomaly  Formula}
Further evaluation of (\ref{4df.5}) is straightforward, but we need to
treat many terms.
Here we introduce a graphical method to express those terms.
Recently it has been shown that
the graphical representation is practically useful to treat  invariants
and covariants in general relativity\cite{SI}.
Because the connectivity of suffixes is visually expressed,
it is very easy to discriminate between independent terms and dependent ones.
Here we apply the technique to the present case:\
to represent global SO($n$) ($n=4$\ in the present
case) covariants and invariants.
We define  the following graphical representation
for $\pl_\al\pl_\be W_\mn\ ,\ \pl_\al N_\m$.
\vs 5
\begin{center}
Fig.1
\end{center}

All independent terms
which could appear in (\ref{4df.5}) are graphically listed up
in App.C.
They are those terms which satisfy the following conditions:\
1)\ Invariants with respect to the global $SO(n)$~ ($n=4$ in this section)
trnasformation of the coordinate;\
2)\ Dimension of (Mass)$^4$;\
3)\ They are composed only of $\pl\pl W$\ , $\pl N$\ and $M$~.
Totally 26 terms appear.
The final evaluation of (\ref{4df.5}) is given by
\begin{eqnarray}
G_2(x,x;t)|_{t^0}^{(\pl\pl W,\pl N,M)}
=\frac{1}{(4\pi)^2}
\{ \frac{1}{45}\pl_\si\pl_\la W_\mn\cdot\pl_\si\pl_\n W_{\m\la}+\cdots
\mbox{(see Table 1)}\q \}\com
                                                            \label{4df.5b}
\end{eqnarray}
and by Table 1 where
the coefficients for all independent terms ,except the overall factor
$1/(4\pi)^2$, are listed.

\vspace{0.5cm}
\begin{tabular}{|c|c|c||c|c|c|}
\hline
Graph      & Expression  & Coeff.  & Graph & Expression & Coeff.          \\
\hline
$\bar {A1}$ & $\pl_\si\pl_\la W_\mn\cdot\pl_\si\pl_\n W_{\m\la}$ & 1/45
& $E1$      & $\pl_\m\pl_\la W_{\la\n}\cdot\pl_\m N_\n$          & 1/12     \\
$\bar {A2}$ & $\pl_\si\pl_\la W_{\la\m}\cdot\pl_\si\pl_\n W_{\mn}$ & -2/45
& $E2$      & $\pl_\m\pl_\la W_{\la\n}\cdot\pl_\n N_\m$          & 1/12     \\
$\bar {A3}$ & $\pl_\si\pl_\la W_{\la\m}\cdot\pl_\m\pl_\n W_{\n\si}$ & -2/45
& $E3$      & $\pl_\m\pl_\n W_{\la\la}\cdot\pl_\n N_\m$          & 0    \\
$\bar {B1}$ & $\pl_\n\pl_\la W_{\si\si}\cdot\pl_\la\pl_\m W_{\mn}$ & -1/90
& $E4$      & $\pl^2 W_\mn\cdot\pl_\n N_\m$                      & 0     \\
$\bar {B2}$ & $\pl^2 W_{\la\n}\cdot\pl_\la\pl_\m W_{\mn}$        & 1/180
& $\bar QR$ & $\pl_\m\pl_\n W_{\mn}\cdot\pl_\la N_\la$           & -1/6     \\
$\bar {B3}$ & $\pl_\m\pl_\n W_{\la\si}\cdot\pl_\m\pl_\n W_{\ls}$ & 1/180
& $\bar PR$ & $\pl^2 W_{\m\m}\cdot\pl_\n N_\n$                   & 1/24     \\
$\bar {B4}$ & $\pl_\m\pl_\n W_{\la\si}\cdot\pl_\la\pl_\si W_{\mn}$ & 1/180
& $F1$      & $\pl_\m N_\n\cdot\pl_\m N_\n$                      & -1/24     \\
${\bar Q}^2$ & $(\pl_\m\pl_\n W_{\mn})^2$                        & 1/18
& $F2$      & $\pl_\m N_\n\cdot\pl_\n N_\m$                      & -1/24     \\
$\bar {C1}$ & $\pl_\m\pl_\n W_{\la\la}\cdot\pl_\m\pl_\n W_{\si\si}$ & 1/360
& $RR$      & $(\pl_\m N_\m)^2$                                   & 1/8     \\
$\bar {C2}$ & $\pl^2 W_{\mn}\cdot\pl^2 W_{\mn}$                   & 1/144
& $M\bar P$ & $M\cdot\pl^2 W_{\m\m}$                              & -1/12    \\
$\bar {C3}$ & $\pl_\m\pl_\n W_{\la\la}\cdot\pl^2 W_{\mn}$         & -1/90
& $M\bar Q$ & $M\cdot\pl_\m\pl_\n W_{\mn}$                       & 1/3    \\
${\bar P}{\bar Q}$ & $\pl^2 W_{\la\la}\cdot\pl_\m\pl_\n W_{\mn}$  & -1/36
& $MR$ & $M\cdot\pl_\m N_\m$                                      & -1/2    \\
${\bar P}^2$ & $(\pl^2 W_{\la\la})^2$                             & 1/288
& $MM$ & $M\cdot M$                                               & 1/2    \\
\hline
\multicolumn{6}{c}{\q}                                                 \\
\multicolumn{6}{c}{Table 1\ \  Anomaly Formula for $(\pl\pl W,\pl N,M)^2$-part
                               of $G_2(x,x;t)|_{t^0}$ }\\
\multicolumn{6}{c}{\q\q\q The overall factor is $1/(4\pi)^2$.
Graph names are defined in App.C.     }\\
\end{tabular}

\vs 1

The result (\ref{4df.3}) for $G_1$\ and the result of Table 1 for
$G_2$\ constitute the anomaly formula in 4 dim.
\vspace{0.5cm}
\subsection{Weyl Anomaly in 4 Dimension}
We apply the formula (\ref{4df.3}) and Table 1  to the Weyl anomaly
calculation of the present example (\ref{f.1}) with $n=4$.
Here we introduce another graphical representation
for the terms appearing in the weak-gravity expansion:\ $g_\mn=\del_\mn
+h_\mn$. We represent\
$\pl_\m\pl_\n h_\ab$ as follows.
\vs 5
\begin{center}
Fig.2
\end{center}
Then we can obtain the following graphical relations from (\ref{f.10b}).
\vs 5
\begin{eqnarray}
\mbox{}                                                     \label{4df.6}
\end{eqnarray}
\vs 5

We focus on $(\pl\pl h)^2$-terms in the anomaly because this type terms
come only from $G_1(x,x;t)|_{t^0}$\ and $(\pl\pl W,\pl N,M)$-part of
$G_2(x,x;t)|_{t^0}$.
All possible terms that could appear in the Weyl anomaly,
are graphically listed up in App.B.
Totally 13 independent terms appear as listed in Table 2.
Inserting the above expressions(\ref{4df.6})
into $(\pl\pl W,\pl N,M)^2$-part of
$G_2(x,x;t)|_{t^0}$(Table 1) we obtain $G_2$-Coeff. of Table 2, where
the overall factor $1/(4\pi)^2$~is omitted. Inserting
$h^2$-part of ($W_\mn,N_\la,M$), defined in (\ref{f.10b}),
into $G_1(x,x;t)|_{t^0}$(\ref{4df.3}) and picking up $(\pl\pl h)^2$-terms,
we obtain $G_1$-Coeff. of Table 2.

\vspace{0.5cm}
\begin{tabular}{|c|c|c|c|c|}
\hline
Graph      & Expression  & $G_2$-Coeff.  & $G_1$-Coeff. & $G_1+G_2$    \\
\hline
$A1 $ & $\pl_\si\pl_\la h_\mn\cdot\pl_\si\pl_\n h_{\m\la}$   &
          $1/45$         & $-7/180$   & $-1/60$          \\
$A2 $ & $\pl_\si\pl_\la h_{\la\m}\cdot\pl_\si\pl_\n h_{\mn}$ &
      $-1/45\cdot 8$     & $-1/90$   & $-1/72$  \\
$A3 $ & $\pl_\si\pl_\la h_{\la\m}\cdot\pl_\m\pl_\n h_{\n\si}$ &
      $-1/45\cdot 8$     & $0$               & $-1/360$ \\
$B1 $ & $\pl_\n\pl_\la h_{\si\si}\cdot\pl_\la\pl_\m h_{\mn}$ &
      $-1/90$            & $2/3\cdot 24$     & $1/60$   \\
$B2 $ & $\pl^2 h_{\la\n}\cdot\pl_\la\pl_\m h_{\mn}$        &
      $ 1/180$           & $-4/15\cdot 24$   & $-1/180$  \\
$B3 $ & $\pl_\m\pl_\n h_{\la\si}\cdot\pl_\m\pl_\n h_{\ls}$ &
      $ 1/180$           & $1/5\cdot 24$     &  $-1/72$  \\
$B4 $ & $\pl_\m\pl_\n h_{\la\si}\cdot\pl_\la\pl_\si h_{\mn}$ &
      $ 1/180$           &  $0$              &  $1/180$  \\
$Q^2$ & $(\pl_\m\pl_\n h_{\mn})^2$                        &
      $ 0$               &   $0$              & $0$  \\
$C1 $ & $\pl_\m\pl_\n h_{\la\la}\cdot\pl_\m\pl_\n h_{\si\si}$ &
  $ 1/360$               &  $-1/144$          & $-1/240$  \\
$C2 $ & $\pl^2 h_{\mn}\cdot\pl^2 h_{\mn}$                   &
      $ 1/144$           &  $-1/15\cdot 24$   & $1/240$  \\
$C3 $ & $\pl_\m\pl_\n h_{\la\la}\cdot\pl^2 h_{\mn}$         &
      $ -1/90$           &  $1/3\cdot 24$     & $1/360$  \\
$PQ $ & $\pl^2 h_{\la\la}\cdot\pl_\m\pl_\n h_{\mn}$  &
      $ 0$               &   $0$              & $0$ \\
$P^2$ & $(\pl^2 h_{\la\la})^2$                       &
      $ 0$               &  $0$               & $0$  \\
\hline
\multicolumn{5}{c}{\q}                                                 \\
\multicolumn{5}{c}{Table 2\ \  Weyl Anomaly of 4 Dim Gravity-Scalar Theory:\
                               $(\pl\pl h)^2$-part }\\
\multicolumn{5}{c}{\qq The overall factor is $1/(4\pi)^2$. Graph names are
defined in App.B.  }\\
\end{tabular}
\vspace{0.5cm}

Now we have evaluated the $(\pl\pl h)^2$-part of the Weyl anomaly completely.
They are expressed by the $(\pl\pl h)^2$-part of the following
invariant quantities.
\begin{eqnarray}
\mbox{Weyl Anomaly}\q =\frac{1}{(4\pi)^2}
\sqg (\al_1 \na^2R+\be_1 R^2+\be_2 R_\mn R^\mn
+\be_3 R_{\mn\ls}R^{\mn\ls})\com\q\nn\\
\al_1=-\frac{1}{180}\com\q \be_1=0\com\q
\be_2=-\frac{1}{180}\com\q \be_3=\frac{1}{180}
                                                  \pr \label{4df.7}
\end{eqnarray}
The above coefficients are obtained by the weak gravity expansion
of the right-hand side and equating the result with that of Table 2.
Weak expansion of all invariants of the right-hand side
is provided in Table 4 of App.B.
{}From the general coordinate transformation invariance, the result
(\ref{4df.7}) is correct to all orders with respect to $h_\mn$.

Here we comment on the reason why the present formulae are
sufficient to determine the anomaly terms completely.
In $n=4$~ dim case, the number of independent (general coordinate)
invariants
which could appear in the Weyl anomaly is 4 as shown in (\ref{4df.7}).
Whereas the number of independent (global SO(4)) invariants of $(\pl\pl h)^2$
\ is 13. Since each term of $(\pl\pl h)^2$~ gives an independent
constraint, $(\pl\pl h)^2$-terms are sufficient to detrmine the Weyl anomaly.
Generally, in n-dim, the number of independent terms which appear in the
Weyl anomaly is far less than that of $(\pl\pl h)^{n/2}$-terms.
Hence the formulae up to $G_{n/2}$~ is sufficient. Furthermore we need not
all terms contained in $G_i\ :\ i=1,2,\cdots ,\frac{n}{2}$.
In $n=4$~ dim case, we have obtained only $(\pl\pl W,\pl N,M)^2$-type
of $G_2$. Other types do not contribute to $(\pl\pl h)^2$-terms.
This is the same situation in the general n-dim case.
Although we expalin for the case of the Weyl anomaly, it is valid
for other anomalies.
\vs 1
\section{Weyl Anomaly in  Fermion-Gravity System}
In the fermion-gravity system, we must treat the vielbein field instead of
the metric field. We explain the weak field expansion
and the graphical representation in this case, by taking an example of
the Weyl anomaly calculation for the system of the
Dirac field coupled to gravity in n dimension.
The  Lagrangian  is given by
\begin{eqnarray}
\Lcal[g_\mn,\psi]=\half \sqg {\bar \psi}i\overnab \psi,
\label{d.1}
\end{eqnarray}
where the notations and conventions are defined by Appendix A.
This Lagrangian is invariant under the following local Weyl transformation:
%
\begin{eqnarray}
& g^\mn(x)'=e^{2\alpha(x)}g^\mn(x)\com\q & \nn\\
& \psi(x)'=\exp\{\frac{n-1}{2}\alpha(x)\}~\psi(x)\com\q
{\bar\psi}(x)'=\exp\{\frac{n-1}{2}\alpha(x)\}~{\bar\psi}(x)\com &
\label{d.2}
\end{eqnarray}
where the notation is the same as Eq.~(\ref{f.2}).

We  redefine the fields as
%
\begin{eqnarray}
{\tilde\psi} \equiv g^{1/4} \psi, \qq
{\tilde{\bar\psi}} \equiv g^{1/4} {\bar\psi},
\label{d.3}
\end{eqnarray}
for the path integral measures to be the general coordinate
invariant\cite{F83}.
In terms of the redefined fields, the Weyl transformation is given by
\begin{eqnarray}
{\tilde\psi}(x)'=e^{-\alpha(x)/2}{\tilde\psi}(x)\com\q
{\tilde{\bar\psi}}(x)'=e^{-\alpha(x)/2}{\tilde{\bar\psi}}(x).
\end{eqnarray}
Then the Jacobian of the above Weyl transformation is written as
%
\begin{eqnarray}
J_{Dirac} &\equiv&
\det\frac{(\pl\tilpsi(y), \pl\tbpsi(y))}{(\pl\tilpsi'(x), \pl\tbpsi'(x))}
=\det~(e^{\al(x)}\del^n(x-y)~)\nn\\
&=& \exp({\rm Tr}~[\al(x)\del^n(x-y)]+O(\al^2))\pr\label{d.4}
\end{eqnarray}
We regularize the delta function $\del^n(x-y)$ as
in the scalar field case.
%
\begin{eqnarray}
G(x,y;t)\equiv <x|e^{-t\Dvec}|y>\com\q t>0\com\nn\\
{\Dvec}_x\equiv - (\fg\nabslash\nabslash\invfg)_x. \label{d.5}
\end{eqnarray}
Then we can express the Weyl anomaly as
\begin{eqnarray}
\ln J_{Dirac} = \lim_{t \rightarrow +0} {\rm Tr}[ \alpha(x) G(x,
y; t)]+O(\al^2)\ .
\end{eqnarray}
Note that the trace contains contraction of the fermion indices.

Since the  action is described by the vielbein, we take the
following weak-field expansion of the vielbein:
\begin{equation}
e_\mu{}^a = \delta_\mu{}^a + f_\mu{}^a\com \label{d.6}
\end{equation}
to solve the differential equation for $G(x, y; t)$~ perturbatively.
The spin connection and the curvature tensor is expanded as
\begin{eqnarray}
\omega_\mu{}^{ab} &=& \half ( - \partial_\mu f_\nu{}^a \delta_\nu{}^b
+ \partial_\nu f_\mu{}^a \delta_\nu{}^b + \partial_\nu f_{\lambda c }
\delta_{\mu c} \delta_\lambda{}^a \delta_\nu{}^b
- ( a \leftrightarrow b)) + O(f^2), \nn \\
R^\lambda{}_{\rho\mu\nu} &=&
  \half \partial_\mu \partial_\rho f_{\lambda a} \delta_{\nu a}
+ \half \partial_\mu \partial_\rho f_{\nu a} \delta_{\lambda a}
- \half \partial_\mu \partial_\lambda f_{\nu a} \delta_{\rho a}
- \half \partial_\mu \partial_\lambda f_{\rho a} \delta_{\nu a} \nn \\
&&- ( \mu \leftrightarrow \nu ) + O(f^2). \label{d.7}
\end{eqnarray}
{}From the above expression, we  obtain $W_\mn$, $N_\mu$ and
$M$ ,in the expanded form, as
%
\begin{eqnarray}
W_{\mu\nu} &=& g^\mn - \delta^\mn
	= - f_{\mu a} \delta_{\nu a} - f_{\nu a} \delta_{\mu a} + O(f^2),\nn \\
N_\lambda &=& - g^\mn \Gamma^\lambda{}_\mn + 2 g^{\mu\lambda} \omega_\mu
         -g^{\m\la}\Ga_{\m\n}^{\n}\nn \\
&=& - ( \partial_\mu f_{\mu a} \delta_{\lambda a}
    + \partial_\mu f_{\lambda a} \delta_{\mu a}
 )
    + \omega_\lambda{}^{ab(1)} \osigma^{ab} + O (f^2), \nn \\
M &=& - g^\mn (\Gamma^\lambda{}_\mn \omega_\lambda
+\Ga_{\m\la}^\la \om_\n)
      + g^\mn (\omega_\mu \omega_\nu +  \partial_\mu \omega_\nu)
      + \half \sigma^{\mu\nu} \sigma^{\lambda\rho} R_{\mu\nu\lambda\rho}
\nn \\
&\qq& +g^\mn(-\half \pl_\m\Ga_{\n\la}^\la+\fourth\Ga_{\m\la}^\la\Ga_{\n\si}^\si
+\half \Ga_\mn^\la \Ga_{\la\si}^\si)                            \nn\\
&=& \delta_\mu{}^c \delta_\nu{}^d \osigma^{cd} \osigma^{ab} \partial_\mu
\omega_\nu{}^{ab(1)} + \half \partial_\mu \omega_\mu{}^{ab(1)} \osigma^{ab}
-\half \pl^2f_\m^{~a}\del_{\m a}+ O(f^2),
\label{d.8}
\end{eqnarray}
where
%
\begin{eqnarray}
\sigma^{\mu\nu} & = & e^\mu{}_a e^\nu{}_b \osigma^{ab}, \nn \\
\omega_\mu{}^{ab(1)} &=& \half ( - \partial_\mu f_\nu{}^a \delta_\nu{}^b
+ \partial_\nu f_\mu{}^a \delta_\nu{}^b + \partial_\nu f_{\lambda c}
\delta_{\mu c} \delta_\lambda{}^a \delta_\nu{}^b
- ( a \leftrightarrow b)).
\label{d.9}
\end{eqnarray}

Here we introduce the graphical representation for $\pl_\m\pl_\n f_\al^{~a}$~
and $\del_{\m a}$ as in Fig.3a and 3b respectively.
\vs 5
\begin{center}
Fig.3
\end{center}
{}From the relation $h_\mn=f_\m^{~a}\del_{\n a}+(\m\change\n)+O(f^2)$~, we
obtain a relation between Fig.2 and Fig.3a, as shown in Fig.4a. In Fig.4b
the 'anti-symmetrized' partner of Fig.4a
(we call it 'anti-symmetric metric ')
is  newly defined.
\vs 8
\begin{center}
Fig.4
\end{center}
Graphically $\pl\pl W,~\pl N$~ and $M$~ are expressed as
\vs{15}
\begin{flushright}
Fig.5
\end{flushright}
\vs{5}
\vs{5}
In Fig.5, $I$~is the $n\times n$~unit matrix.

In order to obtain the Weyl anomaly in 4 dim,
we substitute the above expressions for $n=4$~to the 4 dim anomaly formula of
$G_1(x, x; t)|_{t^0}$ and $G_2(x, x; t)|_{t^0}$. We notice
the anti-symmetric metric looks to appear generally.
In the torsion-less theory of the present example, we know
those terms cancel each other and disappear. For  purely practical reason,
we can take the following gauge
for the external gravitational filed.
\begin{eqnarray}
& f_\m^{~a}\del_{\n a}=f_\n^{~a}\del_{\m a}& \nn\\
\mbox{graphical rep. of gauge}                       \label{d.9b}\\
\nn\\
\nn\\
\nn\\
\nn\\
\nn\\
\nn\\
\nn
\end{eqnarray}

Now the perturbation is done by the power of
$f_\mn\equiv f_\m^{~a}\del_{\n a}+f_\n^{~a}\del_{\m a}$~
which is related with $h_\mn$~as
\begin{eqnarray}
h_\mn=f_\mn+f_\m^{~a}f_{\n a}\com\q
f_\mn\equiv f_\m^{~a}\del_{\n a}+f_\n^{~a}\del_{\m a}\pr
                      \label{d.9c}
\end{eqnarray}

$(\pl\pl f)^2$-terms are obtained
by substituting $f$-part of ($W_\mn,
N_\m,M$), (\ref{d.8}), into $G_2(x,x;t)|_{t^0}$ and
by substituting $f^2$-part of ($W_\mn,
N_\m,M$) into $G_1(x,x;t)|_{t^0}$. Each contribution is separately
listed in Table 3. Since $f_\mn$~ has the same symmetry
as $h_\mn$~,with respect to the suffix-exchange, all possible terms are
the same as $(\pl\pl h)^2$.

\vspace{0.5cm}
\begin{tabular}{|c|c|c|c|c|}
\hline
Graph      & Expression  & $G_2$-Coeff.  & $G_1$-Coeff. & $G_1+G_2$    \\
\hline
$A1' $ & $\pl_\si\pl_\la f_\mn\cdot\pl_\si\pl_\n f_{\m\la}$   &
          $4/45$         & $-1/30$           & $1/18$          \\
$A2' $ & $\pl_\si\pl_\la f_{\la\m}\cdot\pl_\si\pl_\n f_{\mn}$ &
      $-1/90$            & $1/20$            & $7/180$         \\
$A3' $ & $\pl_\si\pl_\la f_{\la\m}\cdot\pl_\m\pl_\n f_{\n\si}$ &
      $-1/90$            & $0$               & $-1/90$         \\
$B1' $ & $\pl_\n\pl_\la f_{\si\si}\cdot\pl_\la\pl_\m f_{\mn}$ &
      $-2/45$            & $0$               & $-2/45$         \\
$B2' $ & $\pl^2 f_{\la\n}\cdot\pl_\la\pl_\m f_{\mn}$        &
      $ 1/45$            & $1/20$            & $13/180$        \\
$B3' $ & $\pl_\m\pl_\n f_{\la\si}\cdot\pl_\m\pl_\n f_{\ls}$ &
      $ 23/360$          & $-1/10$           & $-13/360$       \\
$B4' $ & $\pl_\m\pl_\n f_{\la\si}\cdot\pl_\la\pl_\si f_{\mn}$ &
      $ -7/360$          &  $0$              & $-7/360$        \\
$Q^{2'}$ & $(\pl_\m\pl_\n f_{\mn})^2$                        &
      $ 1/72$            &   $0$             & $1/72$          \\
$C1' $ & $\pl_\m\pl_\n f_{\la\la}\cdot\pl_\m\pl_\n f_{\si\si}$ &
  $ 1/90$                &  $0$           & $1/90$          \\
$C2' $ & $\pl^2 f_{\mn}\cdot\pl^2 f_{\mn}$                   &
      $ 1/36$            &  $-1/20$          & $-1/45$         \\
$C3' $ & $\pl_\m\pl_\n f_{\la\la}\cdot\pl^2 f_{\mn}$         &
      $ -2/45$           &  $0$              & $-2/45$         \\
$PQ' $ & $\pl^2 f_{\la\la}\cdot\pl_\m\pl_\n f_{\mn}$  &
      $ -1/36$           &   $0$             & $-1/36$         \\
$P^{2'}$ & $(\pl^2 f_{\la\la})^2$                       &
      $ 1/72$            &  $0$              & $1/72$          \\
\hline
\multicolumn{5}{c}{\q}                                                 \\
\multicolumn{5}{c}{Table 3\ \  Weyl Anomaly of 4 Dim Gravity-Fermion Theory:\
                               $(\pl\pl f)^2$-part }\\
\multicolumn{5}{c}{\qq The overall factor is $1/(4\pi)^2$. Graph names are
defined in App.B.  }\\
\end{tabular}
\vspace{0.5cm}

In order to express the above result covariantly, we need $(\pl\pl f)^2$-part
of the 4 invariants. As for $R^2,\ R_\mn R^\mn$~ and $R_{\mn\ls}R^{\mn\ls}$,
the expansion coefficients are the same as those given in Table 4.
As for $\na^2R$~, however, the difference between the $f_\mn$-perturbation
and $h_\mn$-perturbation appears due to the existence of the linear term
of $h_\mn$~ in $\na^2R$.
\begin{eqnarray}
\na^2R=\pl^2\pl^2h-\pl^2\pl_\al\pl_\be h_\ab+O(h^2)\com\q
g_\mn=\del_\mn+h_\mn\pr
                      \label{d.9d}
\end{eqnarray}
$O(h^2)$-part gives the same expansion coefficients
as given in Table 4. The $h_\mn$-linear
terms give additional $(\pl\pl f)^2$-terms due to the $f^2$-term in
(\ref{d.9c}). So the second column of Table 4, for the
$(\pl\pl f)^2$-coefficients, should be replaced by
\begin{eqnarray}
\na^2R|_{(\pl\pl f)^2}=(A1')+2(A2')-2(B1')+2(B2')-\frac{3}{2}(B3')\nn\\
+\half (C1')-(C2')-(C3') \
+\{\ -\half (A1')-\half (A2')-\half (B2')+(B3')+\half (C2')\ \}\ ,
                      \label{d.9e}
\end{eqnarray}
where the bracket part ($\{ \cdots\}$) comes from the $h_\mn$-linear terms.
Finally the above result of Table 3 can be written
by the following invariant quantities:
\begin{eqnarray}
\mbox{Weyl Anomaly}\ &\equiv&
\frac{\del}{\del\al(x)}~\ln~J_{Dirac}|_{\al=0}   \nn\\
&=& \sqg (\al_1 \na^2R+\be_1 R^2+\be_2 R_\mn R^\mn
+\be_3 R_{\mn\ls}R^{\mn\ls})\com \label{d.10}
\end{eqnarray}
with the following coefficients:
%
\begin{eqnarray}
\al_1=\frac{1}{30}\com\q \be_1=\frac{1}{72}\com\q
\be_2=-\frac{1}{45}\com\q \be_3=-\frac{7}{360}.\label{d.11}
\end{eqnarray}
This is the same result as the other calculation of the Weyl
anomaly of four-dimensional Dirac field\cite{DC}.

\vs 1
\section{Weyl,
Gravitational and Local Lorentz Anomaly in 2 Dim
Fermion-Gravity System}

$4k+2$ dimensional Weyl fermion coupled to gravity can be anomalous
with respect to the local Lorentz and general coordinate
transformation\cite{AW}. The Lagrangian is expressed by
%
\begin{eqnarray}
\Lcal[g_\mn,\psi]=\sqg \half {\bar \psi}i \overnab
\left( \frac{1-\gamma_5}{2} \right) \psi.
\label{(w.1)}
\end{eqnarray}
The local Lorentz and general coordinate transformation of the
fermion fields are following:
%
\begin{eqnarray}
\delL(\al_{ab})\tilpsi &=& - \half \alpha_{ab} \osigma^{ab} {\tilpsi}, \nn \\
\delG(\xi^\m)\tilpsi  &=& -\half \tilpsi \nabla_\mu \xi^\mu
- \xi^\mu \natil_\mu \tilpsi + \xi^\mu \omega_\mu \tilpsi,, \nn \\
\natil_\m &=& \fg \na_\m\invfg=\fg (\pl_\m+\om_\m)\invfg
                                                   \com\label{w.2}
\end{eqnarray}
where $\tilpsi$ is defined in Eq.(\ref{d.3}), and
$\alpha_{ab}$ is a local Lorentz transformation gauge parameter,
$\xi^\mu$ is a general coordinate transformation parameter and
$\om_\m=\half\osigma^{ab}\om_{\m ab}$.
We take the following
transformation instead of $\delG$.
%
\begin{equation}
\delc\tilpsi = [\delG(\xi^\mu) + \delL(\xi^\mu \omega_{\mu ab})]\tilpsi
             = -\half \tilpsi \nabla_\mu \xi^\mu
                  - \xi^\mu \natil_\mu \tilpsi \pr    \label{w.3}
\end{equation}
We define the following operators:
%
\begin{eqnarray}
\Dslash_{L} \equiv \fg \nabslash \invfg \left(
\frac{1-\gamma_5}{2} \right), \qq
\Dslash_{L}{}^{\dag} \equiv \fg \nabslash \invfg
\left( \frac{1 + \gamma_5}{2} \right),\label{w.4}
\end{eqnarray}
Then the Lagrangian is written as
%
\begin{eqnarray}
\Lcal[g_\mn,\psi]= \half {\tilde{\bar \psi}}i
{\overleftrightarrow\Dslash}_{L}{\tilde\psi}. \label{w.5}
\end{eqnarray}

In this section, we explicitly calculate 2 dim local Lorentz and
general coordinate anomaly using the anomaly formulae.

\vspace{0.5cm}

\subsection{Local Lorentz Anomaly in 2 Dim Weyl Fermion}

We write the Jacobian of local Lorentz transformation of measure
${\cal D}\tbpsi{\cal D}\tilpsi$ as $J_{LL}$.
The Jacobian is written as
%
\begin{eqnarray}
\ln J_{LL} &=& (\ln J_{LL})_{\tilde\psi}
+ (\ln J_{LL})_{\tilde{\bar\psi}} \nn \\
&=& \half {\rm Tr}_{\tilde\psi} \left[
\alpha_{ab}(x) \osigma_{ab} \delta^2(x-y) \right]
- \half {\rm Tr}_{\tilde{\bar\psi}}
\left[ \alpha_{ab}(x) \osigma_{ab}
\delta^2(x-y) \right].\label{w.6}
\end{eqnarray}
Because $\Dslash_L$~is not hermitian, we have some freedom in
the choice of $\Dvec_x$\cite{F85}.
We take it in such a way that the covariant anomaly is obtained.
So we regularize the delta functions in (\ref{w.6}) as
%
\begin{eqnarray}
\ln J_{LL} &=& \lim_{t \to +0} \left( \half \right)
{\rm Tr} \left[
\alpha_{ab}(x) \osigma_{ab}
<x| \exp(t \Dslash_{L}{}^{\dag} \Dslash_{L} )|y>  \right] \nn \\
&&- \lim_{t \to +0} \left( \half \right) {\rm Tr} \left[
\alpha_{ab}(x) \osigma_{ab}
<x| \exp(t \Dslash_{L} \Dslash_{L}{}^{\dag} )|y>
\right]. \label{w.7}
\end{eqnarray}
(\ref{w.7}) is rewritten as
\begin{eqnarray}
\ln J_{LL} &=& \lim_{t \to +0} \left( \half \right)
{\rm Tr} \left[
\alpha_{ab}(x) \osigma_{ab}
<x| \left( \frac{1+\gamma_5}{2} \right)
\exp(t\fg\nabslash\nabslash\invfg)|y>  \right] \nn \\
&&- \lim_{t \to +0} \left( \half \right) {\rm Tr} \left[
\alpha_{ab}(x) \osigma_{ab}
<x| \left( \frac{1-\gamma_5}{2} \right) \exp(t\fg\nabslash\nabslash\invfg)|y>
\right] \nn \\
&=& \lim_{t \to +0} \left( \half \right) {\rm Tr} \left[
 \alpha_{ab}(x) \osigma_{ab} \gamma_5
<x| \exp(t\fg\nabslash\nabslash\invfg)|y>  \right].
\label{w.8}
\end{eqnarray}
%
\begin{eqnarray}
G(x, y; t) = <x|\exp(t\fg\nabslash\nabslash\invfg)|y>\com\nn\\
\Dvec_x = - (\fg\nabslash\nabslash\invfg)_x,\label{w.9}
\end{eqnarray}
which is the same as n dim Dirac fermion case (\ref{d.5}).
The expressions for $W_{\mn}(x)$, $N_\mu(x)$ and $M(x)$ are
(\ref{d.8}).

To calculate $G(x, x; t) = <x|\exp(t\fg\nabslash\nabslash\invfg)|x> $,
we use the 2 dim
formula (\ref{2df.2}) in Sec.3 and substitute
(\ref{d.8}) to (\ref{2df.2}). Then we obtain
%
\begin{eqnarray}
G_1( x, x; t) &=& \frac{1}{4\pi} \biggl[ \frac{1}{6} \partial^2
(f_{\mu a} \delta_{\mu a})
- \frac{2}{3} \partial_\mu \partial_\nu (f_{\mu a} \delta_{\nu a} )
+ \half \partial_\lambda ( 2 \partial_\mu f_{\mu a} \delta_{\lambda a}
- \partial_\lambda f_{\mu a} \delta_{\mu a} )
\nonumber \\
&&+ \delta_{\mu c} \delta_{\nu d} \osigma^{cd} \osigma^{ab}
\partial_\mu \omega_{\nu ab}{}^{(1)} \biggr] + O(f^2),\label{w.10}
\end{eqnarray}
where $\omega_{\nu ab}{}^{(1)}$ is defined by Eq.~(\ref{d.9}).
Since
$\osigma^{ab} = \frac{i}{2} \gamma_5 \epsilon^{ab}$
is valid in 2 dim and the trace of
the odd number products of $\gamma_5$ vanishes,
the anomaly term is calculated as
%
\begin{eqnarray}
A_{LL} &\equiv&\al_{ab}(x)\frac{\pl}{\pl\al_{ab}(x)}~\ln~J_{LL}\nn\\
&=& \frac{1}{4 \pi} \alpha_{ab} \epsilon^{ab} \frac{i}{12}
\left[ \frac{1}{3} \partial^2 f_{\mu a} \delta_{\mu a}
- \frac{1}{3} \partial_\mu \partial_\nu f_{\mu a} \delta_{\nu a} \right]
+ O(f^2) \nonumber \\
&=& \alpha_{ab} \frac{1}{4 \pi} \frac{i}{24} \epsilon^{ab}
 \sqrt{g} R \pr
\label{w.11}
\end{eqnarray}
This reproduces the result of \cite{AW,FTY}.


\subsection{General Coordinate Anomaly in 2 Dim Weyl Fermion}

The calculation of the general coordinate anomaly is slightly technical.
Let us calculate the gravitational anomaly for the
transformation $\delc$.
Compared with all other transformations in the present text,
only the general coordinate one has
the (covariant) derivative term.
The Jacobian of the measure ${\cal D}\tbpsi{\cal D}\tilpsi$
with respect to the transformation $\delc$ denoted as $J_{cov}$
is written as
%
\begin{eqnarray}
\ln J_{cov} &=& (\ln J_{cov})_{\tilde\psi}
+ (\ln J_{cov})_{\tilde{\bar\psi}} \nn \\
&=& \half {\rm Tr}_{\tilde\psi} \left[
 \left( \half {\nabla}_\mu \xi^\mu(x) + \xi^\mu(x)
{\natil}_\mu \right) \delta^2(x-y) \right] \nn \\
&&+ \half {\rm Tr}_{\tilde{\bar\psi}}
\left[ \left( \half {\nabla}_\mu \xi^\mu(x)
+ {\tilde {\leftnabla}}_\mu \xi^\mu(x) \right)
\delta^2(x-y) \right].
\label{w.12}
\end{eqnarray}
We can regularize the Jacobian as
%
\begin{eqnarray}
\ln J_{cov} &=& \lim_{t \to +0} {\rm Tr} \biggl[
<x| \left( \half {\nabla}_\mu \xi^\mu(x) +
\xi^\mu(x) {\natil}_\mu \right) \exp(t \Dslash_{L}{}^{\dag}
\Dslash_{L})|y> \nn \\
&&+ <x| \left( \half {\nabla}_\mu \xi^\mu(x)
+ {\tilde {\leftnabla}}_\mu \xi^\mu(x) \right) \exp(t \Dslash_{L}
\Dslash_{L}{}^{\dag})|y> \biggr] \nn \\
&=& \lim_{t \to +0} {\rm Tr} \biggl[
<x| \left( \half {\nabla}_\mu \xi^\mu(x) +
\xi^\mu(x) {\natil}_\mu \right) \left( \frac{1+\gamma_5}{2}
\right) \exp(t\fg\nabslash\nabslash\invfg)_y|y> \nn \\
&&+ <x| \left( \half {\nabla}_\mu \xi^\mu(x)
+ {\tilde {\leftnabla}}_\mu \xi^\mu(x) \right) \left
( \frac{1-\gamma_5}{2} \right)
\exp(t\fg\nabslash\nabslash\invfg)|y> \biggr] \nn \\
&=& \lim_{t \to +0} \left(\half \right) {\rm Tr}
\left[ \gamma_5 <x| \fg ( \xi^\mu(x) \nabla_\mu
+ \nabla_\mu \xi^\mu(x) )\invfg\exp(t\fg\nabslash\nabslash\invfg)|y>
                                                        \right]\nn \\
&=& \lim_{t \to +0} \left(\half \right) {\rm Tr}
\left.\left[ \frac{1}{t} \gamma_5 <x|\exp\{t\fg ( \nabslash\nabslash
+ \xi^\mu \nabla_\mu  + \nabla_\mu \xi^\mu )\invfg\}_|y> \right]
\right|_{\xi^1}.
\label{w.13}
\end{eqnarray}
In the last expression, taking the linear term of $\xi$~ is understood.
This trick was taken in \cite{F85}.
So we consider the following heat-kernel.
%
\begin{eqnarray}
G(x, y; t) = <x|\exp\{t\fg ( \nabslash\nabslash
+ \xi^\mu \nabla_\mu  + \nabla_\mu \xi^\mu )\invfg\}|y> , \nn \\
\Dvec_x = - \left( \fg (\nabslash\nabslash + \xi^\mu \nabla_\mu  +
\nabla_\mu \xi^\mu)\invfg \right)_x .
\label{w.14}
\end{eqnarray}
Then  $W_{\mu\nu}$, $N_\mu$ and
$M$~ , corresponding to (\ref{w.14}), is given by, up to
the linear order of $f_{\mu a}$,
%
\begin{eqnarray}
W_{\mu\nu} &=& g^\mn - \delta^\mn
	= - f^{\mu a} \delta^{\nu a} - f^{\nu a} \delta^{\mu a} + O(f^2),\nn \\
N^\lambda &=& - g^\mn \Gamma^\lambda{}_\mn - g^{\mu\lambda}
\Gamma^\nu{}_{\mu\nu} + 2 g^{\mu\lambda} \omega_\mu
	+ 2\xi^\lambda \nn \\
&=& - ( \partial_\mu f_{\mu a} \delta_{\lambda a}
    + \partial_\mu f_{\lambda a} \delta_{\mu a} )
    + \omega_\lambda{}^{ab(1)} \osigma^{ab} + 2 \xi_\lambda + O (f^2), \nn \\
M &=& - g^\mn (\Gamma^\lambda{}_\mn \omega_\lambda
      + \Gamma^\lambda{}_{\mu\lambda} \omega_\nu )
      +g^\mn(-\half \pl_\m\Ga_{\n\la}^\la+\fourth\Ga_{\m\la}^\la\Ga_{\n\si}^\si
+\half \Ga_\mn^\la \Ga_{\la\si}^\si)                            \nn\\
&&      + \half \sigma^{\mu\nu} \sigma^{\lambda\rho} R_{\mu\nu\lambda\rho}
      + g^\mn \omega_\mu \omega_\nu + g^\mn \partial_\mu \omega_\nu
      + 2 \xi^\mu \omega_\mu + \partial_\mu \xi^\mu
                       \nn \\
&=& \delta_\mu{}^c \delta_\nu{}^d \osigma^{cd} \osigma^{ab} \partial_\mu
\omega_\nu{}^{ab(1)}
+ \half \partial_\mu \omega_\mu{}^{ab(1)} \osigma^{ab}
- \half \partial^2 f_\mu{}^a \delta_{\mu a}
+ \xi_\mu \omega_\mu{}^{ab(1)} \osigma^{ab}
+ \partial_\mu \xi_\mu \nn \\
&& -\xi^\m\pl_\m f^{~a}_\n\del_{a\n} + O(f^2) . \label{w.15}
\end{eqnarray}
Since the last expression of (\ref{w.13}) has the factor $t^{-1}$,
and has the expansion variable $\xi^\m$~in addition to $f_{\m a}$,
we need two dim formula of
$G_1 (x, y; t)|_{t^1}$ and $G_2 (x, y; t)|_{t^1}$ to
obtain the finite term at $t \to +0$.

{}From the formulae (\ref{f.17}) and (\ref{f.18}), we note
that
%
\begin{eqnarray}
\frac{1}{4 \pi t} G^{(4)}_l(x, y; t) |_{t^0} = G^{(2)}_l(x, y;
t) |_{t^{1}}, \qquad \mbox{for} \quad l = 1, 2, \cdots,
\label{w.20}
\end{eqnarray}
where $G^{(n)}_l(x, y; t)$ is $G_l(x, y; t)$ in $n$ dimension.
Therefore,
from the result (\ref{4df.3}) in
Sec.4, we obtain the two dimensional formula of
$G_1 (x, x; t)|_{t^1}$ as
%
\begin{eqnarray}
G_1 (x, x; t)|_{t^1} &=& \frac{t}{4 \pi} \biggl[ - \frac{1}{120}
(\partial^2)^2 W_{\mu\mu}(x) + \frac{1}{20} \partial^2 \partial_\mu
\partial_\nu W_{\mu\nu}(x) \nn \\
&&- \frac{1}{12} \partial^2 \partial_\mu
N_\mu(x) + \frac{1}{6} \partial^2 M(x) \biggr], \label{w.16}
\end{eqnarray}
This term produces the only total divergence term, so it does not
contribute to the gravitational anomaly.
Next we calculate the contribution from $G_2 (x, y; t)|_{t^1}$.
Since the trace of the odd number products of $\gamma_5$ vanishes and we
need the terms proportional to ${(\xi_\mu{})^1}$, the necessary terms in
$G_2 (x, y; t)|_{t^1}$ are $W_\mn$-independent terms. They are given by,
from (\ref{4df.4b}),
%
\begin{eqnarray}
G_2 (x, x; t)|_{t^1} = \frac{t}{4 \pi} \biggl[ - \frac{1}{12}
(\partial^2 N_\mu(x)) N_\mu(x)
- \frac{1}{24} \partial_\mu N_\nu(x) \partial_\mu N_\nu(x) \nn \\
- \frac{1}{24} \partial_\nu N_\mu(x) \partial_\mu N_\nu(x)
 + \frac{1}{8} \partial_\mu N_\mu(x) \partial_\nu N_\nu(x)
- \frac{1}{2} \partial_\mu N_\mu(x) M(x)
+ \frac{1}{2} M(x) M(x) \biggr], \label{w.17}
\end{eqnarray}
Substituting the expression (\ref{w.15}) to (\ref{w.17}), we obtain
%
\begin{eqnarray}
G_2 (x, x; t)|_{t^1} = \frac{t}{4 \pi} \biggl[
- \frac{1}{6} \xi_\mu \partial^2 \omega_\mu{}^{ab} \osigma^{ab}
+ \frac{1}{6} \xi_\mu \partial_\mu \partial_\nu \omega_\nu{}^{ab}
\osigma^{ab} \biggr] + \mbox{tot. div} + O(f^2) , \label{w.18}
\end{eqnarray}
[tot. div] is the total divergence terms, which does not produce
the gravitational anomaly. From (\ref{w.13}) and (\ref{w.18}),
we obtain the following result:
%
\begin{eqnarray}
A_{cov}(x)&\equiv&\xi_\m(x)\frac{\pl}{\pl\xi_\m(x)}\ln~J_{cov}\nn\\
&=& \frac{1}{4 \pi} \frac{i}{12} \xi_\mu \epsilon_{\nu\lambda}
[\partial^2 \partial_\nu f_{\mu a} \delta_{\lambda a}
+ \partial^2 \partial_\nu f_{\lambda a} \delta_{\mu a} \nn \\
&& - \partial_\mu \partial_\rho \partial_\lambda f_{\rho a} \delta_{\nu a}
- \partial_\mu \partial_\rho \partial_\nu f_{\lambda a}
\delta_{\rho a}] + O(f^2) \nn \\
&=& \frac{1}{4 \pi} \frac{i}{12} \xi_\mu \sqrt{g}
e^{\nu\lambda} \nabla^\rho R^\mu{}_{\rho\nu\lambda}     \nn \\
&=& - \frac{1}{4 \pi} \frac{i}{12} \xi_\mu \sqrt{g}
e^{\mu\nu} \nabla_\nu R ,
\label{w.19}
\end{eqnarray}
which reproduces the known result about the gravitational
anomaly\cite{AW,FTY}.

The relation (\ref{w.20}) suggests that
two dimensional gravitational anomaly is related to the four
dimensional certain anomaly. It is the chiral anomaly. Their
relations will be closely investigated in Sec.7.2.


\subsection{Relations of Anomalies}

In this subsection and in 7.2, we examine  relations among some
anomalies.
Let us consider the Weyl transformation of two-dimensional Dirac
field. The Lagrangian is (\ref{d.1}) and the Weyl transformation
is the same as Eq.~(\ref{d.2}). So we obtain the Weyl anomaly of
two dimensional Dirac field coupled to gravity as follows:
\begin{eqnarray}
\ln J^{2d}_{Dirac} = \lim_{t \rightarrow +0} {\rm Tr}[ \alpha(x)
 <x|e^{-t\Dvec}|y>]\pr\nn\\
G(x,y;t)\equiv <x|e^{-t\Dvec}|y>\com\q t>0\com\label{rel.1}\\
{\Dvec}_x\equiv - (\fg\nabslash\nabslash\invfg)_x\pr\nn
\end{eqnarray}
Because  $\osigma^{ab} = \frac{i}{2} \gamma_5
\epsilon^{ab}$\ is valid in 2 dim,
the local Lorentz anomaly (\ref{w.8}) becomes
\begin{eqnarray}
\ln J_{LL} &=& \lim_{t \to +0} \left( \frac{i}{4} \right) {\rm Tr} \left[
 \alpha_{ab}(x) \epsilon_{ab}
<x|\exp(t\fg\nabslash\nabslash\invfg)|y>
\right]\pr
                                             \label{rel.2}
\end{eqnarray}
By defining
\begin{eqnarray}
K(x) &=& \lim_{t \to +0} \left.{\rm tr} \left[
 <x|\exp(t\fg\nabslash\nabslash\invfg)|y>
\right]\right|_{x=y}\com                    \label{rel.3}
\end{eqnarray}
we can express the local Lorentz
anomaly and the Weyl anomaly as
\begin{eqnarray}
\ln J_{LL} &=& \intx\frac{i}{4} \alpha_{ab}(x) \epsilon_{ab} K(x), \nn \\
\ln J^{2d}_{Dirac} &=& \intx\alpha(x) K(x).
\end{eqnarray}
This relation is due to the simplicity of 2 dim. It is not valid for
higher dimensions.
{}From the above result the Weyl anomaly of two dimensional Dirac field
is obtained by
replacing $\frac{i}{4} \alpha_{ab}(x) \epsilon_{ab}$  by
$\alpha(x)$ in two dimensional local Lorentz anomaly (\ref{w.11}).
So we can obtain the Weyl anomaly of the Dirac fields as
\begin{eqnarray}
\ln J^{2d}_{Dirac}
= \intx~\alpha(x) \frac{1}{4 \pi} \frac{1}{6} \sqrt{g} R\pr
\end{eqnarray}

Next, we note the relation of the local Lorentz anomaly
and general coordinate anomaly pointed out by \cite{FTY}.
When we set $A^{ab}(x)$ and $A_\mu(x)$ as follows:
\begin{eqnarray}
& A_{LL}(x) = \alpha_{ab}(x) A^{ab}(x)\com\q
 A_{cov}(x) = \xi^\mu(x) A_\mu(x)\com
\end{eqnarray}
then the following relation is satisfied:
\begin{eqnarray}
\nabla^\mu [e_\mu{}^a(x) e_\nu{}^b(x) A^{ab}(x)] = - \half A_\nu(x).
\end{eqnarray}
This relation can be assured in this case explicitly.
It is valid for higher dimensions.

\vs{0.5}
In Sec.5 of ref.\cite{AW}, the 2 dim gravitational anomaly was calculated
in the light-cone gauge for the external gravity.
The counter-term is treated and
the Weyl anomaly is obtained in a
different way from the present one. The present approach does not
fix the gauge, which is desirable because the gravity is not
quantized.
\section{Chiral Anomaly in Flat and Gravitational Theories}
\vspace{0.5cm}
\subsection{Chiral U(1) Anomaly in 4 Dim QED}
We take a simple model of Euclidean QED.
\begin{eqnarray}
&\Lcal={\bar \psi}i\ga^aD_a\psi-m{\bar \psi}\psi-\fourth F_{ab}F^{ab}\com&
                                                 \nn\\
&F_{ab}=\pl_bA_a-\pl_aA_b\com\q D_a=\pl_a+ieA_a\com\q
a=1,\cdots,4  \com& \label{chi.1}
\end{eqnarray}
where $\psi$\ and $A_a$\ are Dirac fermion and the gauge(photon) field
respectively.
The gamma matrices $\ga^a$~ in this subsection are the same as $\ogamma^a$~
defined in App.A.  The chiral U(1) transformation is given by
\begin{eqnarray}
\psi'=e^{i\al(x)\ga_5}\psi\com\q
{\bar \psi}'={\bar \psi}~e^{i\al(x)\ga_5}
                                                 \com \label{chi.2}
\end{eqnarray}
Then the anomaly measure and its regularization are given as
\begin{eqnarray}
\ln~J_{chiral}^{QED}=\ln~\frac{\pl(\psi'(x),{\bar \psi}'(x))}
{\pl(\psi(y),{\bar \psi}(y))}=
{}~2i~\mbox{Tr}\{\al(x)\ga_5\del^4(x-y)\}+O(\al^2)\com\nn\\
I_4\del^4(x-y)=\lim_{t\ra +0}G(x,y;t)\com\q
G(x,y;t)\equiv <x|e^{-t(-D^2)}|y>\pr \label{chi.3}
\end{eqnarray}
The ingredients for the anomaly formula (\ref{4df.4b}) are given by
\begin{eqnarray}
&(\frac{\pl}{\pl t}+(-D^2))~G(x,y;t)=0\com&\nn\\
&-D^2\equiv -D_aD_a=-\pl_a\pl_a-V\com\q
V=N_a\pl_a+M\com  &                   \label{chi.4}\\
&N_a=2ieA_a\com\q
M=-e^2A_aA_a+ie\pl_aA_a-\frac{ie}{4}[\ga^a,\ga^b]F_{ab}\pr &\nn
\end{eqnarray}
{}From the anomaly formula (\ref{4df.4b}), the chiral anomaly is given by
\begin{eqnarray}
\half \left.\frac{d}{d\al(x)}~\ln~J^{QED}_{chiral}\right|_{\al=0}
&=&~\mbox{tr}~i\ga_5G_2^{flat}(x,x;t)|_{t^0}
=\frac{1}{(4\pi)^2}\mbox{tr}(i\ga_5\half M^2)\nn\\
&=&\frac{i}{(4\pi)^2}\frac{e^2}{2}F_{ab}{\tilde F}^{ab}\com\q
{\tilde F}^{ab}\equiv \ep^{abcd}F_{cd}
                                                   \label{chi.5}
\end{eqnarray}

\subsection{Chiral U(1) Anomaly in 4 Dim Fermion-Gravity System
and Relation to 2 Dim Gravitational Anomaly}

In this subsection, we consider the 4 dim
gravitational chiral $U(1)$ anomaly
\cite{Kim,DS,EF,ET,Gip,Alv}
 and its connection to two
dimensional gravitational anomaly in the context of our
formalism.
The Lagrangian (\ref{d.1}) is classically invariant under the
following gravitational chiral transformation of the Dirac
fermion :
%
\begin{eqnarray}
{\tilde\psi}(x)'=e^{i\alpha(x)\gamma_5}{\tilde\psi}(x)\com\q
{\tilde{\bar\psi}}(x)'={\tilde{\bar\psi}}(x)e^{i\alpha(x)\gamma_5}.
\label{gch.1}
\end{eqnarray}
%
The chiral anomaly is computed by the same procedure as
in Sec.2 and Sec.5 and is given by
%
\begin{eqnarray}
\ln J_{chiral} = -2i \lim_{t \rightarrow +0} {\rm Tr}[ \alpha(x)
 \gamma_5 G(x, y; t)],
\label{gch.2}
\end{eqnarray}
where $G(x, y; t)$~ is given by (\ref{d.5}).
%
%
We do not explain the detailed calculation because it is the same
procedure as in the previous sections.
We obtain the chiral anomaly as
%
\begin{eqnarray}
\ln J_{chiral} &=& - \frac{1}{(4 \pi)^2} \frac{i}{12}\intfx \alpha(x)
\epsilon^{\lambda\rho\sigma\tau}
( \partial_\lambda \partial_\nu f_{\mu a} \delta_{\rho a}
+ \partial_\lambda \partial_\nu f_{\rho a} \delta_{\mu a} ) \nn \\
&& \times [ \partial_\sigma \partial_\nu f_{\mu b} \delta_{\tau b}
+ \partial_\sigma \partial_\nu f_{\tau b} \delta_{\mu b}
- ( \mu \leftrightarrow \nu ) ] + O(f^2) \nn \\
&=& - \frac{1}{(4 \pi)^2} \frac{i}{24} \intfx \alpha(x)
\sqrt{g} e^{\lambda\rho\sigma\tau} R_{\mu\nu\lambda\rho}
R^{\mu\nu}{}_{\sigma\tau} ,
\label{gch.4}
\end{eqnarray}
where
%
$ e^{\lambda\rho\sigma\tau} = g^{-1/2}
\epsilon^{\lambda\rho\sigma\tau}, $
%
and $\epsilon^{\lambda\rho\sigma\tau}$ is the antisymmetric
unit tensor in 4 dim.


We investigate the relation between 2 dim gravitational
anomaly and 4 dim chiral anomaly suggested in
(\ref{w.20})\cite{F85,AW}.
In order to reduce 4 dim quantities to 2 dim ones, we divide
4 dim expression to $2+2$ dim expression.
We can express 4 dim
gamma matrices by 2 dim ones:
%
\begin{eqnarray}
{\gamf^1} & =  {\gamt^1} \otimes \sigma^3\ ,\
{\gamf^2}  =  {\gamt^2} \otimes \sigma^3\ ,\
{\gamf^3}  =  1 \otimes \sigma^1\ ,\
{\gamf^4}  =  1 \otimes \sigma^2\ ,\nn \\
{\gamf_5} & =  - i {\gamt^1} {\gamt^2}
\otimes \sigma^3
= {\gamt_5} \otimes \sigma^3,
                                             \label{gch.8}
\end{eqnarray}
where $Q_{(4)}$ is 4 dim quantity and $Q_{(2)}$ is 2
dim one and $\sigma^i$ are the Pauli matrices.
We consider the special background gravitational field as in the
Kaluza-Klein type dimensional reduction.
Let us take the case that the spin connection
$\omega_\mu{}^{ab}$ is given by the following equation:
%
\begin{eqnarray}
{\omef}_\mu{}^{ab}(x) = \left\{
\begin{array}{@{\ }ll}
\omet_\mu{}^{ab}(\xhat)    & \mbox{if $\mu = 1, 2$ and $a,b = 1, 2$,} \\
- i \xi_\mu(\xhat)         & \mbox{if $\mu = 1, 2$ and $a=3$, $b=4$,} \\
i \xi_\mu(\xhat)           & \mbox{if $\mu = 1, 2$ and $a=4$, $b=3$,} \\
0 & \mbox{otherwise,}
\end{array}
\right.
\label{omegadef}\\
(x)=(x^1,\cdots,x^4)\com\q (\xhat)=(x^1,x^2)\pr\nn\\
\end{eqnarray}
Using the relations
\begin{eqnarray}
&& \sigf{}^{ab} = \sigt^{ab}  \otimes 1 \quad \mbox{ if $a, b =
1, 2$}, \qquad \sigf{}^{34} = \half 1 \otimes i \sigma^3, \nn \\
&& \sigf{}^{43} = - \half 1 \otimes i \sigma^3,  \qquad \mbox{etc.},
\end{eqnarray}
we obtain
%
\begin{eqnarray}
\omef_\mu{} = \omet_\mu{}^{bc} ( \sigt^{bc} \otimes 1 )
+ \xi_\mu (1 \otimes \sigma^3)\q\mbox{for }\m=1,2\com\nn\\
\omef_3{} =\omef_4{} =0\pr
                                        \label{simgaexp}
\end{eqnarray}
It is sufficient that we consider the correspondence of the operators
$\Dvec_x$ in two theories.
So we expand $\Dvec_{(4)x} = (\fg\nabslash \nabslash\invfg)_{(4)}$
to $2 + 2$
dimensional expression
under Eq.(\ref{omegadef}) as
%
\begin{eqnarray}
(\nabslash \nabslash)_{(4)}  =
\left( \sum_{\mu =1}^2 \gamt^{\mu} \na_\mu{}_{(2)} \right)
\left( \sum_{\nu =1}^2 \gamt^{\nu} \na_\nu{}_{(2)} \right)
\otimes 1
+ 1 \otimes \left( \sum_{\mu =3}^4 \partial_\mu{}^2 \right) \nn \\
+ \left( \sum_{\mu =1}^2 \gamt^{\mu} \na_\mu{}_{(2)} \right)
\left( \sum_{\nu =1}^2 \gamt^{\nu} \xi_\nu \right)
\otimes \sigma^3
+ \left( \sum_{\mu =1}^2 \gamt^{\mu} \xi_\mu \right)
\left( \sum_{\nu =1}^2 \gamt^{\nu} \na_\nu{}_{(2)} \right)
\otimes \sigma^3
 \nn \\
 + \left( \sum_{\mu =1}^2 \gamt^{\mu} \xi_\mu \right)
\left( \sum_{\nu =1}^2 \gamt^{\nu} \xi_\nu \right)
\otimes 1,
\label{nablaexp}
\end{eqnarray}

For simplicity, we make a calculation by the symbolical
expression
$G(x,y;t)\equiv$\linebreak
$<x|e^{-t\Dvec}|y>$
of the solution in the heat equation.
If we define
%
\begin{eqnarray}
\Achi(x,y;t) \equiv {\rm tr}[  \gamf_5 G(x, y; t)],
\label{gch.6}
\end{eqnarray}
then the chiral anomaly is expressed as
%
\begin{eqnarray}
\ln J_{chiral} = \lim_{t\ra +0}\intfx\{-2i \alpha(x) \Achi(x,x;t)\}.
\label{gch.7}
\end{eqnarray}
Thus,from (\ref{nablaexp}), we can express $\Achi(x,y;t)$ as
\begin{eqnarray}
\Achi(x,y;t) &=&  {\rm tr} [(\gamt_5{} \otimes \sigma^3)
<x|\exp \{~ t~\fg [ 1 \otimes \sum_{\mu = 3}^{4} \partial_\mu{}^2  \nn \\
&& + ( \nabslat  \nabslat
+ \xislash\xislash ) \otimes 1 + ( \nabslat \xislash
+ \xislash  \nabslat ) \otimes \sigma^3
]\invfg\} |y>].
\end{eqnarray}
where $\nabslat=\gamt^\m\na_{(2)\m},
\xislash=\gamt^\mu \xi_\mu$.
Now we focus on the first-order in
$\xi^\mu$ .
%
\begin{eqnarray}
\Achi(x,y;t)|_{(\xi^\mu)^1} = t\times {\rm tr_{(2)}}
[(\gamt_5{} <x| \fg (\nabslat \xislash
 + \xislash \nabslat )\invfg
\nn \\
 \times  \exp [t\fg \nabslat  \nabslat\invfg ] |y>]
\otimes {\rm tr_{(2)}}<x| \exp [ t \sum_{\mu = 3}^{4}
\partial_\mu{}^2 ] |y>\ .
\label{relaAB}
\end{eqnarray}
Using the relation
\begin{eqnarray}
{\rm Tr_{(2)}}<x| \exp [ t \sum_{\mu = 3}^{4}
\partial_\mu{}^2 ] |y>={\rm tr}(\frac{1}{4\pi t}I_2) = \frac{1}{2 \pi t}\com
\end{eqnarray}
and the expression for $A_{cov}(x)$, (\ref{w.13}),
%
the equation (\ref{relaAB}) says
\begin{eqnarray}
\lim_{t\ra +0}\Achi(x,x;t)|_{(\xi^\mu)^1} = \frac{1}{2 \pi}\times 2A_{cov}(x),
\end{eqnarray}
for the background gravitational field (\ref{omegadef}).

Using this relation, we can calculate gravitational anomaly $A_{cov}(x)$
from the chiral anomaly $\Achi(x,x;t)$.
For (\ref{omegadef}),
the curvature tensor becomes
\begin{eqnarray}
R_{(4)\mu\nu}{}^{ab} = \left\{
\begin{array}{@{\ }ll}
R_{(2)\mu\nu}{}^{ab}     & \mbox{if $\mu,\nu = 1, 2$ and $a,b = 1, 2$,} \\
( - i \nabla_\mu \xi_\nu + i \nabla_\nu \xi_\mu )
                         & \mbox{if $\mu,\nu = 1, 2$ and $a=3$, $b=4$,} \\
( i \nabla_\mu \xi_\nu - i \nabla_\nu \xi_\mu )
                         & \mbox{if $\mu,\nu = 1, 2$ and $a=4$, $b=3$,} \\
0                        & \mbox{otherwise}
\end{array}
\right.
\end{eqnarray}
If we substitute the above expression into $\Achi(x,x;t)$-part of
(\ref{gch.4}), we obtain
\begin{eqnarray}
A_{cov}(x) & = &  \pi\lim_{t\ra +0} \Achi(x,x;t)|_{(\xi^\mu)^1} \nn \\
&= & \frac{1}{4 \pi} \frac{1}{48} \epsilon^{ab34}
R_{(2)}{}^{\mu\nu}{}_{ab} ( - i \nabla_\mu \xi_\nu + i \nabla_\nu \xi_\mu )
\times 2 \nn \\
& = & -\frac{1}{4 \pi} \frac{i}{12} \nabla_\mu \xi_\nu
R_{(2)}{}^{\mu\nu}{}_{ab} \epsilon^{ab} \nn \\
& = & -\frac{1}{4 \pi} \frac{i}{12}\sqg e^{\nu\mu} \xi_\nu \nabla_\mu R
+ [\mbox{tot. div}],
\end{eqnarray}
where the [tot. div] is the total divergence term.
This term does not contribute the gravitational anomaly, so it
turns out to give the same result as in the Sect. 6.2.

Generally, we can extend above discussion to the higher dimension.
The following relations are satisfied in the solutions,
$G_l^{(n)}(x, y; t)$, of the heat equation:
\begin{eqnarray}
\frac{1}{4 \pi t} G_l^{(4k)}(x, y; t) |_{t^m} = G_l^{(4k-2)}(x,
y; t) |_{t^{m+1}},
\end{eqnarray}
whence $4k-2$ dimensional gravitational anomaly is related to $4k$
dimensional chiral anomaly.
We can connect two anomalies by the similar discussion in this
section.

\vs 2
As demonstrated in Sect.6 and 7, the presented anomaly formulae provide
a powerful tool to calculate all kinds of anomalies concretely.
This is complementary to ref.\cite{AW} where the chiral  anomaly
is derived from the result of the differential geometry and the gravitational
anomaly is indirectly derived using the relation between the two anomalies.

\section{Discussions}
\vspace{0.5cm}
We have presented a systematic approach to general anomaly calculation.
It is based on the propagator theory.
Fujikawa's general standpoint about anomaly is taken. In the evaluation
of the Jacobian due to the change of the path-integral measure,
we take the heat-kernel regularization.
The 2 dim and 4 dim general anomaly formulae are obtained.
The known various anomalies and some relations between them are
explicitly  derived from the formulae.
We have also presented the graphical
technique for the representation of (global SO(n)) invariants and
covariants appearing in the weak field expansion calculation.
For the flat space(-time) case, the 4 dim anomaly
formula reduces to the 'tHooft's
1-loop counter-term formula except  total derivative terms.

\vs 1
We make some additional comments and refer to the future directions.
\begin{enumerate}
\item
The present systematic calculation and the graph technique
enable us to do the anomaly
calculation using a computer. It opens the possibility to calculate
anomalies in other interesting theories such as Weyl anomaly
of the gravitational theories, in  higher dimensions,
coupled with various matter fields.
\item
The anomaly calculations so far are mainly concerned with the matter
(1-loop) quantum effect. The effect due to gauge or gravity quantum
excitation is also important, especially for the Weyl anomaly\cite{KPZ,FT}.
When
the present approach is combined with the background field method, matter
fields and gauge or gravity fields can be equally treated.
The generalization to such a direction is interesting.
\item
The higher-loop effects are  important for the Weyl anomaly,
because it is directly related to the renormalization-group $\be$-function.
The higher-loop generalization of the present approach is also interesting.
\item
Recently (4 dim) super Yang-Mills theory is vigorously investigated
in relation to the dual symmetry\cite{SW1,SW2}. In the problem some anomalies,
such as Konishi anomaly\cite{Koni}
and the super-Weyl anomaly, play an important
role. We hope the present approach helps to
clarify some part of this interesting problem.
\end{enumerate}

\vs 1
Renormalization and anomaly are two outstandingly important aspects
of quantum field theory. They have been giving us various rich information
and helping us to understand the field theory.
The direct relation between
the 1-loop counter-term formula and the anomaly formula
clearly shows the intimate relation between them.

\vs 2
\begin{flushleft}
{\bf Acknowledgement}
\end{flushleft}
The authors thank K. Fujikawa and R. Endo
for discussions and comments about the present work. They express
gratitude to N. Nakanishi and K. Fujikawa
for reading the manuscript carefully. This work has been completed
during the stay of one of the authors (S.I.) at RIMS, Kyoto Univ..
He thanks all its members for hospitality.

\vs 2
\begin{flushleft}
{\Large\bf Appendix A.\ Notations and Some Useful Formulae}
\end{flushleft}
Here we collect the present notation and some useful formulae.

\begin{flushleft}
{\bf A.1\ Basic Notation}
\end{flushleft}

\begin{eqnarray}
\Ga^\la_\mn=\half g^\ls (\pl_\m g_{\si\n}+\pl_\n g_{\si\m}-\pl_\si g_\mn)\com
R^\la_{~\m\n\si}=\pl_\n\Ga^\la_{\m\si}+\Ga^\la_{\tau\n}\Ga^\tau_{\m\si}-
\n\change\si\com\nn\\
R_\mn =R^\la_{~\mn\la}\com\q R=g^\mn R_\mn\com\q g=+\mbox{det}g_\mn\pr
                                                      \label{not.1}
\end{eqnarray}
The covariant derivative for a vector field $A_\mu$:
\begin{eqnarray}
\nabla_\mu A_\nu = \partial_\mu A_\nu - \Gamma^\lambda{}_{\mu\nu}
A_\lambda\pr
\end{eqnarray}
The covariant derivative for general tensor field
$T_{\Sigma} = T^{\sigma_1\cdots\sigma_r}_{\tau_1\cdots\tau_s}$:
\begin{eqnarray}
\nabla_\mu T_{\Sigma} = \partial_\mu T_{\Sigma} +
\Gamma^\lambda{}_{\mu\nu} [T_{\Sigma}]^\nu{}_\lambda,
\end{eqnarray}
where
\begin{eqnarray}
[T^{\sigma_1\cdots\sigma_r}_{\tau_1\cdots\tau_s}]^\nu{}_\lambda
= \sum_{p=1}^r \delta^{\sigma_p}{}_\lambda
T^{\sigma_1\cdots\sigma_{p-1}\nu\sigma_{p+1}\cdots\sigma_r}
_{\tau_1\cdots\tau_s}
- \sum_{q = 1}^s \delta^\nu{}_{\tau_q}
T^{\sigma_1\cdots\sigma_r}_{\tau_1\cdots\tau_{q-1}\lambda\tau_{q+1}\cdots\tau_s}.
\end{eqnarray}

\begin{flushleft}
{\bf A.2\ Weak-Field Expansion}
\end{flushleft}

\begin{eqnarray}
g_\mn &=& \del_\mn+h_\mn\com\nn\\
R_{\mn\ls} &=& \half (\pl_\la\pl_\n h_{\m\si}-\pl_\la\pl_\m h_{\n\si}
       -\la\change\si)+O(h^2)\com\nn\\
R_\mn &=& \half (\pl_\m\pl_\n h-\pl_\m\pl_\al h_{\al\n}-\pl_\n\pl_\al h_{\al\m}
+\pl^2 h_\mn )+O(h^2)\com\nn\\
R &=& \pl^2 h-\pl_\al\pl_\be h_\ab +O(h^2)\com\nn\\
\Ga^\la_\mn &=& \half (\pl_\m h_{\la\n}+\pl_\n h_{\la\m}-\pl_\la h_\mn)
                                        +O(h^2)\com\nn\\
\sqg &=& 1+\half h_{\m\m}+O(h^2)\pr                   \label{not.2}
\end{eqnarray}

\begin{flushleft}
{\bf A.3\ Integral Formula}
\end{flushleft}

\begin{eqnarray}
\intnw~ e^{-Aw^2} &=& (\frac{\pi}{A})^{\frac{n}{2}}\com\nn\\
\intnw~ w^\m w^\n e^{-Aw^2} &=&
\frac{\del^\mn}{2}(\frac{\pi}{A})^{\frac{n}{2}}\frac{1}{A}\com\nn\\
\intnw~ w^\m w^\n w^\la w^\si e^{-Aw^2} &=&
\frac{[\mn\ls]}{4}(\frac{\pi}{A})^{\frac{n}{2}}\frac{1}{A^2}\com\nn\\
\intnw~ w^{\m_1}\cdots w^{\m_6} e^{-Aw^2} &=&
\frac{[\m_1\cdots\m_6]}{8}(\frac{\pi}{A})^{\frac{n}{2}}\frac{1}{A^3}\com\nn\\
\intnw~ w^{\m_1}\cdots w^{\m_{2s}} e^{-Aw^2} &=&
\frac{[\m_1\cdots\m_{2s}]}{2^s}(\frac{\pi}{A})^{\frac{n}{2}}\frac{1}{A^s}
                                                      \com\label{not.3}
\end{eqnarray}
where
\begin{eqnarray}
[\mn\ls] &\equiv& \del^\mn\del^\ls+\del^{\m\si}\del^{\n\la}
                               +\del^{\m\la}\del^{\n\si}\com\nn\\
\ [\m_1\cdots\m_6]
 &\equiv& \del^{\mu_1\mu_2}[\mu_3\cdots\mu_6]
+\del^{\mu_1\mu_3}[\cdots]+\cdots+\del^{\mu_1\mu_6}[\cdots]  \com\nn\\
\ [\m_1\cdots\m_{2s}]
 &\equiv& \del^{\mu_1\mu_2}[\mu_3\cdots\mu_{2s}]
+\del^{\mu_1\mu_3}[\cdots]+\cdots+\del^{\mu_1\mu_{2s}}[\cdots]  \pr
                                                      \label{not.4}
\end{eqnarray}

\begin{eqnarray}
G_0(x;t) &=& \frac{1}{(4\pi t)^{n/2}}e^{-\frac{x^2}{4t}}~I_N\com\nn\\
\int d^nx~x^{\m_1}\cdots x^{\m_{2s}}G_0(x;t) &=&
[\m_1\cdots \m_{2s}](2t)^s~I_N\com\nn\\
\int^1_0dr~r^p(1-r)^q &=& \frac{p!~q!}{(p+q+1)!}\pr
                                                      \label{not.5}
\end{eqnarray}

\begin{flushleft}
{\bf A.4\ Spinor and Vielbein}
\end{flushleft}

%
\begin{eqnarray}
&& \{ \ogamma^a, \ogamma^b \} = 2 \delta^{ab}\com
 (\ogamma^a){}^{\dag} = \ogamma^a \com
 \osigma^{ab} = \frac{1}{4} [ \ogamma^a, \ogamma^b ] = - \osigma^{ba}\com
                                                  \nonumber \\
&& \gamma^\mu = e^\mu{}_a \ogamma^a\com
 \{ \gamma^\mu, \gamma^\nu \} = 2 g^{\mn}, \nn \\
&& {\bar\psi} = \psi^{\dag} \gamma^4\pr
\end{eqnarray}
\begin{eqnarray}
&& \omega_\mu{}^a{}_b = ( \Gamma^\lambda{}_{\mu\nu} e_\lambda{}^a
	- \partial_\mu e_\nu{}^a ) e^\nu{}_b\com
 \omega_\mu = \half \osigma^{ab} \omega_{\mu ab}\pr
\end{eqnarray}
%
\begin{eqnarray}
&& \nabla_\mu \psi = (\partial_\mu + \omega_\mu ) \psi\com
 {\bar\psi} {\overleftarrow\nabla}_\mu = {\bar\psi}
({\overleftarrow\partial}_\mu - \omega_\mu ), \nn \\
&& \nabslash = \gamma^\mu \nabla_\mu\com
 {\bar \psi}\overnab \psi = {\bar\psi} ( \gamma^\mu \nabla_\mu \psi ) -
( {\bar \psi} \gamma^\mu {\overleftarrow\nabla}_\mu ) \psi\pr
\end{eqnarray}
%
%
In 4 dim
\begin{eqnarray}
&& \ogamma_5 = - \ogamma^1 \ogamma^2 \ogamma^3 \ogamma^4\com
 (\ogamma_5)^2 = 1, \nn \\
&& e^{\lambda\rho\sigma\tau} = g^{-1/2}
\epsilon^{\lambda\rho\sigma\tau}\com
\epsilon^{1234}=+1\pr
\end{eqnarray}
where $\epsilon^{\lambda\rho\sigma\tau}$ is the antisymmetric
unit tensor in four dimension.

In 2 dim
\begin{eqnarray}
&& \ogamma_5 = - i \ogamma^1 \ogamma^2\com
 (\ogamma_5)^2 = 1\com \nn \\
&& \osigma^{ab} = \frac{i}{2} \ogamma_5 \epsilon^{ab}\com
 e^{\lambda\rho} = g^{-1/2}\epsilon^{\lambda\rho}\com
\epsilon^{12}=+1\pr
\end{eqnarray}

\vs 1
\begin{flushleft}
{\Large\bf Appendix B.\ Graphical Representation of $\pl_\m\pl_\n h_\ab$
and $(\pl\pl h)^2$-Invariants}
\end{flushleft}

In the weak field expansion of  n-dim Euclidean gravity
:\ $g_\mn=\del_\mn+h_\mn,|h|\ll 1$,\
we must generally treat  global SO(n)-tensors
which are composed of $h_\mn,\ \pl_\al h_\mn,$\linebreak
$\pl_\al\pl_\be h_\mn,
\cdots$. Let us focus here on those tensors which are composed only of
$\pl_\al\pl_\be h_\mn$. We introduce its graphical representation
as shown in Fig.2 in subsect.4.4 of the text.
Contraction of suffixes is graphically represented by gluing the dotted
lines with the same suffix. For example

\vs 5
\begin{center}
Fig.6
\end{center}
Independent $\pl\pl h$-scalars (Dimension $(Mass)^2$)
are the following two graphs.
\vs 5
\begin{center}
Fig.7
\end{center}
In the ordinary literal expression, P=$\pl^2 h\ $,\ Q=$\pl_\m\pl_\n h_\mn$\ .
Similarly we can list up all independent $(\pl\pl h)^2$-scalars
(Dimension $(Mass)^4$)
as follows. They are grouped by the number of the suffix-loop.
\begin{flushleft}
{loop no\ =1}
\end{flushleft}
\vs 5
\begin{center}
Fig.8
\end{center}

\begin{flushleft}
{loop no\ =2}
\end{flushleft}
\vs {9.5}

\begin{center}
Fig.9
\end{center}

\begin{flushleft}
{loop no\ =3}
\end{flushleft}
\vs {9.5}
\begin{center}
Fig.10
\end{center}

\begin{flushleft}
{loop no\ =4}
\end{flushleft}
\vs 5
\begin{center}
Fig.11
\end{center}

We have totally 13 invariants (10 connected,\ 3 disconnected).
Their literal expressions are listed in Table 2 of the text(subsect.4.4).

As some simple applications, we give the  weak-gravity expansion
of Riemann tensors
in the graphical way as follows.
\vs 7
\begin{eqnarray}
\mbox{}                                                     \label{appb.1}
\end{eqnarray}
\vs 7

The products of Riemann tensors are similarly expressed,
and their $O((\pl\pl h)^2)$-parts
are the linear combinations of the above 13 invariants.
There are 4 independent invariants of dimension $(Mass)^4$\ including
total derivatives:\
$\na^2R\ ,\ R^2\ ,\ R_\mn R^\mn\ ,\linebreak
R_{\mn\la}R^{\mn\ls}$\ .
The 13 coefficients for each invariants
are listed in Table 4.

\vspace{0.5cm}
\begin{tabular}{|c||c|c|c|c|}
\hline
Graph      & $\qq\na^2R\qq$  & $\qq R^2\qq$  & $\qq R_\mn R^\mn\qq$
                                             &  $R_{\mn\ls}R^{\mn\ls}$  \\
\hline
$A1 $ &    $1$       & $0$            & $0$                 & $-2$       \\
$A2 $ &    $2$       &  $0$           & $\half$             & $0$        \\
$A3 $ &    $0$       &  $0$           & $\half$             & $0$        \\
$B1 $ &    $-2$      &  $0$           & $-1$                & $0$        \\
$B2 $ &    $2$       &  $0$           & $-1$                & $0$        \\
$B3 $ &$-\frac{3}{2}$&  $0$           & $0$                 & $1$        \\
$B4 $ &    $0$       &  $0$           & $0$                 & $1$        \\
$Q^2$ &    $0$       &  $1$           & $0$                 & $0$        \\
$C1 $ & $\half$      &  $0$           & $\fourth$           & $0$        \\
$C2 $ &    $-1$      &  $0$           & $\fourth$           & $0$        \\
$C3 $ &    $-1$      &  $0$           & $\half$             & $0$        \\
$PQ $ &    $0$       &  $-2$          & $0$                 & $0$        \\
$P^2$ &    $0$       &  $1$           & $0$                 & $0$        \\
\hline
\multicolumn{5}{c}{\q}                                                 \\
\multicolumn{5}{c}{Table 4\ \  Weak-Expansion of Invariants with
(Mass)$^4$-Dim.
                              :\  $(\pl\pl h)^2$-Part }\\
\end{tabular}

\vs 1

We use the above result in the Weyl anomaly calculation of Sec.4.4
and Sec.5. The results of this appendix B are valid to general dimension
of space(-time).
\vspace{0.5cm}

\begin{flushleft}
{\Large\bf Appendix C.\ Graphical Representation of Anomaly Formula}
\end{flushleft}

In the anomaly formula in 4 dim (Sect.4.3), we must deal with invariants
made of two tensors out of ($\pl\pl W,\pl N,M$). We introduce a useful
graphical representation for them. First we define it for the basic
elements: $\pl_\al\pl_\be W_\mn$\ and $\pl_\al N_\m$\ as shown
in Fig.1 of the text (Sect 4.3).

We notice the symmetry of $\pl_\al\pl_\be W_\mn$\
, with respect to the exchange of suffixes,
is the same
as that of $\pl_\al\pl_\be h_\mn$\ . Therefore the first group of
invariants $(\pl\pl W)^2$\ are obtained just by the substitution of
a single solid line  by a double solid line in the 13 invariants in App.B.
\begin{flushleft}
i)\ $(\pl\pl W)^2$
\end{flushleft}
\begin{displaymath}
\begin{array}{ccccc}
{\bar {A1}},\q & {\bar {A2}},\q & {\bar {A3}},\q &              &           \\
{\bar {B1}},\q & {\bar {B2}},\q & {\bar {B3}},\q & {\bar {B4}},\q
                                                                &{\bar Q}^2,\\
{\bar {C1}},\q & {\bar {C2}},\q & {\bar {C3}},\q & {\bar P}{\bar Q},\q & \\
{\bar P}^2   &&&&
\end{array}
\end{displaymath}
Other groups of independent invariants are as follows.
\begin{flushleft}
ii)\ $\pl\pl W \times \pl N$
\end{flushleft}

\vs {10}
\begin{center}
Fig.12
\end{center}
\begin{flushleft}
iii)\ $\pl N \times \pl N$
\end{flushleft}

\vs 5
\begin{center}
Fig.13
\end{center}
\begin{flushleft}
iv)\ $M \times (\pl\pl W,\pl N,M)$
\end{flushleft}

\vs {10}
\begin{center}
Fig.14
\end{center}

Totally
there are 16 connected ones and 10 disconnected ones. We use,in the text,
the above notation for 26 independent invariants. The ordinary (literal)
expression for each graph is listed in Table 1 in Sect.4.3.

\newpage
\begin{flushleft}
{\Large\bf Figure Captions}
\end{flushleft}
\begin{itemize}
\item
Fig.1\ Graphical representation of $\pl_\al\pl_\be W_{\mn}$\ and
$\pl_\al N_\m$.
\item
Fig.2\ Graphical representation of $\pl_\m\pl_\n h_{\ab}$.
\item
Fig.3\ Graphical representation of $\pl_\m\pl_\n f_\al^{~a}$\ ((a))and
$\del_{\m a}$\ ((b)).
\item
Fig.4\ (a)~Graphical relation between $\pl_\la\pl_\si h_{\mn}$\ (Fig.2)
and $\pl_\la\pl_\si f_\m^{~a}$\ (Fig.3);\
(b)~Graphical definition of 'anti-symmetric metric'.
\item
Fig.5\ Graphical representation of $\pl_\la\pl_\si W_{\mn},\
\pl_\si N_\la$\ and $M$\ for the fermion-gravity system.
\item
Fig.6\ Graphical representation of $\pl_\al\pl_\be h_{\al\n}$.
\item
Fig.7\ Graphical representation of $\pl^2 h$\ (P) and $\pl_\m\pl_\n h_\mn$\
(Q).
\item
Fig.8\ Graphical representation of $(\pl\pl h)^2$-scalars. The number of
suffix-loops is 1.
\item
Fig.9\ Graphical representation of $(\pl\pl h)^2$-scalars. The number of
suffix-loops is 2.
\item
Fig.10\ Graphical representation of $(\pl\pl h)^2$-scalars. The number of
suffix-loops is 3.
\item
Fig.11\ Graphical representation of $(\pl\pl h)^2$-scalars. The number of
suffix-loops is 4.
\item
Fig.12\ Graphical representation of the anomaly formula\ :\
$\pl\pl W\times \pl N$.
\item
Fig.13\ Graphical representation of the anomaly formula\ :\
$\pl N\times \pl N$.
\item
Fig.14\ Graphical representation of the anomaly formula\ :\
$M\times (\pl\pl W,~\pl N,~M)$.
\end{itemize}


\begin{thebibliography}{99}
\bibitem{foot1} 
{
This approach can be regarded as the anomaly-version of the 'tHoot's
renormalization analysis using the counter-term formula\cite{tH}.
}
\bibitem{foot2} 
{
The case of special $W^{ij}_\mn$~(Riemann geometry) was done by
Gilkey\cite{Gil}, where he analyzed using the normal coordinates in order to
keep the general covariance manifestly.
}
\end{thebibliography}
\end{document}